\documentclass[aps,a4paper,12pt, twocolumns]{revtex4-1}
 \pdfoutput=1 
\usepackage{float}
\usepackage{fancyhdr}
\usepackage{color}
\usepackage{graphicx}
\usepackage{epsfig} 
\usepackage{amssymb}
\usepackage{amsmath,amsfonts,verbatim}
\usepackage{booktabs}
\usepackage{hyperref}
\usepackage{eufrak}
\usepackage{mathrsfs}
\usepackage{upgreek}
\usepackage[footnotesize,singlelinecheck=off,justification=raggedright]{caption}
\def\ket#1{\mathinner{|{#1}\rangle}}

\def\Bra#1{\left<1>}
\def\g{{\rm g}}
\def\myPhi{\Theta}

\DeclareMathAlphabet{\mathbbmsl}{U}{bbm}{m}{sl}
{\catcode`\|=\active\gdef\Braket#1{\left<\mathcode`\|"8000\let|\bravert {#1}\right>}}
\def\bravert{\egroup\,\vrule\,\bgroup}

\def\Tr{\mathop{\mbox{\normalfont Tr}}\nolimits}

\def\CM{{\mathcal M}}
\def\e{{\mathrm e}}
\def\v{{\mathbf v}}
\def\chartheta#1#2{\vartheta\!\left[\begin{smallmatrix}#1\\#2\end{smallmatrix}\right]\!}
\def\hattheta#1#2{\widehat\vartheta\!\left[\begin{smallmatrix}#1\\#2\end{smallmatrix}\right]\!}
\begin{document}
\def\comment#1{}

\title{Entanglement entropy and M\"obius transformations for critical 
fermionic chains.}

 \author{Filiberto Ares\footnote{Corresponding author.}}
 \email{ares@unizar.es}
 \affiliation{Departamento de F\'{\i}sica Te\'orica, Universidad de Zaragoza,
 50009 Zaragoza, Spain}
 \author{Jos\'e G. Esteve}
 \email{esteve@unizar.es}
 \affiliation{Departamento de F\'{\i}sica Te\'orica, Universidad de Zaragoza,
 50009 Zaragoza, Spain}
 \affiliation{Instituto de Biocomputaci\'on y F\'{\i}sica de Sistemas
 Complejos (BIFI), 50009 Zaragoza, Spain}
 \author{Fernando Falceto}
 \email{falceto@unizar.es}
 \affiliation{Departamento de F\'{\i}sica Te\'orica, Universidad de Zaragoza,
 50009 Zaragoza, Spain}
 \affiliation{Instituto de Biocomputaci\'on y F\'{\i}sica de Sistemas
 Complejos (BIFI), 50009 Zaragoza, Spain}

 \author{Amilcar R. de Queiroz}
 \email{amilcarq@gmail.com}
 
 \affiliation{Instituto de Fisica, Universidade de Brasilia, 
 Caixa Postal 04455, 70919-970, Bras\'{\i}lia, DF, Brazil}


\begin{quote}
\end{quote}

\begin{abstract}
  Entanglement entropy may display a striking new symmetry under M\"obius 
transformations. This symmetry was analysed in our previous work for the case of a 
non-critical (gapped) free homogeneous fermionic chain invariant under parity and charge 
conjugation.  In the present work we extend and analyse this new symmetry in several 
directions. First, we show that the above mentioned symmetry also holds when parity 
and charge conjugation invariance are broken. Second we extend this new symmetry to the 
case of critical (gapless) theories. Our results are further supported by numerical 
analysis. For some particular cases, analytical demonstrations show the validity 
of the extended symmetry. We finally discuss the intriguing parallelism of this new 
symmetry and space-time conformal transformations.
\end{abstract}

\maketitle

\section{Introduction}

Entanglement entropy is recognized as a most useful tool for the study of critical 
properties of quantum extended systems \cite{Holzhey, Latorre, CarCal, Li, Hsieh, 
Voloda1, Voloda2, Lepori, Carrasco1, Carrasco2}. For instance, in one dimension, the entanglement 
entropy in the ground state of a critical system grows logarithmically with the size of 
the subsystem. The coefficient of the logarithmic term encodes the central charge of the 
corresponding conformal field theory. In fact, in the pioneer works \cite{Holzhey}, 
\cite{CarCal}, the behaviour under space-time conformal transformations
of the two-point function in a 1+1 dimensional critical theory,
was used to compute the leading scaling behaviour of the
entanglement entropy.

In the case of a general translation invariant free fermionic chain one can enlarge the 
group of symmetries of the entanglement entropy, including not only space-time symmetries
(like the conformal transformations) but also a realization of the M\"obius group that 
acts on the coupling constants of the theory.

The first glimmer of this new symmetry came from Ref. \cite{Franchini}, where Franchini, 
Its, Jin and Korepin discovered that in the space of couplings of the XY spin chain there 
are ellipses and hyperbolas along which the von Neumann entanglement entropy of the ground 
state is constant. They unravelled this property by a direct analysis of the expressions 
for the entropy, obtained by Its, Jin and Korepin \cite{Its} and Peschel \cite{Peschel2} 
in terms of elliptic integrals.

Recently, in Ref. \cite{Ares4}, we noticed that this invariance  occurs in more general 
spinless fermionic chains and it is not only valid for the von Neumann entropy but also 
for the R\'enyi entanglement entropy \cite{Franchini2}. 
This fact implies something deeper: the full spectrum of the two-point correlation 
function remains invariant on these curves. 

In the same paper \cite{Ares4}, we traced back the origin of this symmetry. For any 
fermionic chain described by a quadratic Hamiltonian, the R\'enyi entanglement entropy of 
the ground state can be written in terms of the determinant of the two-point correlation 
matrix. In the case of periodic and homogeneous Hamiltonians, the correlation function of 
an interval of contiguous sites is a block Toeplitz matrix. The asymptotics of a block 
Toeplitz determinant can be obtained by solving a Riemann-Hilbert problem \cite{Widom, 
Deift}. For certain cases, where the model has a mass gap, the solution to this problem 
has been found \cite{Its, Its2}. The determinant can be written in terms of the Riemann 
theta function of a compact Riemann surface, defined by a hyperelliptic curve whose genus 
is related to the range of the couplings. Then, it is easy to see that the determinant of 
the correlation matrix is invariant under the action of M\"obius transformations on the 
hyperelliptic curve. Moreover, since it only depends on the couplings of the theory, 
we have families of gapped theories connected by a subgroup of these transformations with 
the same ground state entanglement entropy. From this point of view, the conics of 
constant entanglement entropy found in \cite{Franchini} correspond to flows of the 
M\"obius transformations for genus one.

In this paper, we further analyse the behaviour of the entanglement entropy under these 
M\"obius transformations. In particular, we shall consider critical theories as well as 
Hamiltonians that break parity (the same as reflection) and/or charge conjugation 
symmetries. These aspects are relevant when we compute the entanglement entropy 
\cite{Kadar, Ares3}. In this sense, recall that even if the Hamiltonian violates parity 
this does not necessarily imply that its vacuum breaks this symmetry. Actually, if the 
Hamiltonian is non critical, the ground state will preserve parity and the entanglement 
entropy will remain invariant under the above M\"obius transformations. On the contrary, 
if the theory is massless, the vacuum can break the parity symmetry. 

From the above geometrical perspective, critical theories with a parity invariant 
vacuum correspond to hyperelliptic curves where some pairs of branch points degenerate at 
the unit circle. Therefore, the associated Riemann surface is pinched and the behaviour of 
the entropy changes. In fact, we shall see that it is no longer invariant  under 
M\"obius transformations, but it changes like a product of homogeneous fields. A different 
case is that of a critical theory with a parity symmetry breaking vacuum. We shall show 
that in this situation the entanglement entropy transforms also like homogeneous fields, 
but of different dimension.

Another natural question is what happens when the subsystem is made out of several 
disjoint intervals. The difficulty here is that the correlation matrix is not block 
Toeplitz anymore, but its principal submatrix. However, using the 
expression proposed in \cite{AEF} for the determinant of these kind of matrices, we are 
able to deduce the transformation law of the entanglement entropy for these disconnected 
subsystems. This reveals a striking similarity with its behaviour under conformal 
transformations on the real space when the theory is massless. 
 
The paper is organized in the following way. In Section II, we introduce the basic 
notations and definitions as well as the models we shall study, discussing its critical 
properties and the symmetries of its ground state. In section III, we focus on 
non-critical theories with ground state being parity invariant. We shall show that the 
entanglement entropy of a single interval is invariant under M\"obius transformations. The 
proof presented here is more general and simpler than that of \cite{Ares4}. Moreover, the 
new approach reveals in a transparent way that the M\"obius symmetry of entanglement 
entropy is a consequence of the asymptotic invariance of the spectrum of the correlation 
matrix.  Section IV is dedicated to critical theories. The first part of this section is 
concerned with theories with parity invariant ground state while in the 
second part we consider those theories with parity symmetry breaking ground state. 
In both cases we find that the transformation law of the entropy is 
analogous to that of a product of homogeneous fields inserted at the discontinuities of 
the symbol of the correlation matrix. Their scaling dimension is the ingredient that 
distinguishes between vacua that preserve parity and those that break it. In Section V, we 
shall extend this transformation law to the case of several disjoint intervals, comparing 
it with the behaviour under global conformal transformations in the real space. Finally, 
in Section VI we shall end with some conclusions. In the Appendices, we analyse in detail 
the case of Hamiltonians with parity and charge conjugation symmetries and study the 
degenerate limit of their associated hyperelliptic curves, both if they correspond to 
massive or massless theories.

\section{Entanglement entropy in the free fermionic chain}
\label{sec:chain}

Let us consider a unidimensional chain of $N$ spinless fermions described by the 
following periodic, quadratic, translation invariant Hamiltonian with long range 
couplings ($L<N/2$) 
\begin{equation}\label{ham}
 H=\frac{1}{2}\sum_{n=1}^N \sum_{l=-L}^L \left(2 A_l a_n^\dagger a_{n+l}
+B_l a_n^\dagger a_{n+l}^\dagger-\overline B_l a_n a_{n+l}\right).
\end{equation}
Fermionic annihilation and creation operators
at site $n$,  $a_n$ and $a_n^\dagger$, satisfy the canonical
anticommutations relations
\begin{equation}
\{a_n, a_m\}=\{a_n^\dagger, a_m^\dagger\}=0, 
\quad \{a_n^\dagger, a_m\}=\delta_{nm},
\end{equation}
and we fix periodic boundary conditions, $a_{n+N}=a_n$.
\comment{We further consider that reflection symmetry ($a_n\mapsto i a_{N-n}$)
      is preserved, i.e. $A_l\in\mathbb{R}$ $\forall l$.}
The Hamiltonian is Hermitian when $A_{-l}=\overline A_l$. In addition, we shall impose, 
without loss of generality, that $B_{-l}=-B_l$. 
\comment{For the moment, let us consider that reflection-charge conjugation ($a_n\mapsto i 
a_{N-n}^\dagger$) can be broken; that is, the pairing terms can be $B_l\in\mathbb{C}$.}

Since the Hamiltonian is translation invariant, we may 
introduce the Fourier modes
$$b_k= \frac1{\sqrt{N}}\sum_{n=1}^N {\rm e 
}^{in\theta_k}a_n,\qquad\theta_k=\frac{2\pi k}{N}.$$ After performing the appropriate 
Bogoliubov transformation $d_k=\cos\xi_k b_k + i \sin\xi_k b^\dagger_{-k}$, 
we obtain the diagonalized Hamiltonian (see e.g. \cite{Ares3}),
\begin{equation}\label{diag-ham}
H=\mathcal{E}+\sum_{k=0}^{N-1}\Lambda(\theta_k) 
\left(d_k^\dagger d_k-\frac{1}{2}\right),
\end{equation}
where
$$\mathcal{E}=\frac{1}{2}\sum_{k=0}^{N-1}
\Theta(\mathrm{e}^{i\theta_k})
$$
is an irrelevant constant shift in the energy levels and, for $\theta\in(-\pi,\pi]$,
\begin{equation}\label{disp-relat-1}
\Lambda(\theta)=\sqrt{
  \Theta^+(\mathrm{e}^{i\theta})^2
  +|\Xi(\mathrm{e}^{i\theta})|^2
} +\Theta^-(\mathrm{e}^{i\theta})
\end{equation}
is the dispersion relation. Both these functions may be expressed in terms of the 
Laurent polynomials
\begin{equation}\label{laurent}
\Theta(z)=\sum_{l=-L}^L A_l z^l,\quad  
\Xi(z)=\sum_{l=-L}^L B_l z^l\quad{\rm and}\quad \Theta^\pm(z)=
\frac{\Theta(z)\pm \Theta(z^{-1})}2,
\end{equation}
that map meromorphically the Riemann sphere $\overline{\mathbb{C}} 
=\mathbb{C}\cup\{\infty\}$ to itself. Due to the properties of the coupling 
constants we have $\overline{\Theta^{\pm}(\overline z)}=\Theta^{\pm}(z^{-1 })=
\pm\Theta^\pm(z)$
and $\Xi(z^{-1})=-\Xi(z)$.

\begin{figure}[h]
  \centering
    \resizebox{16cm}{13cm}{\includegraphics{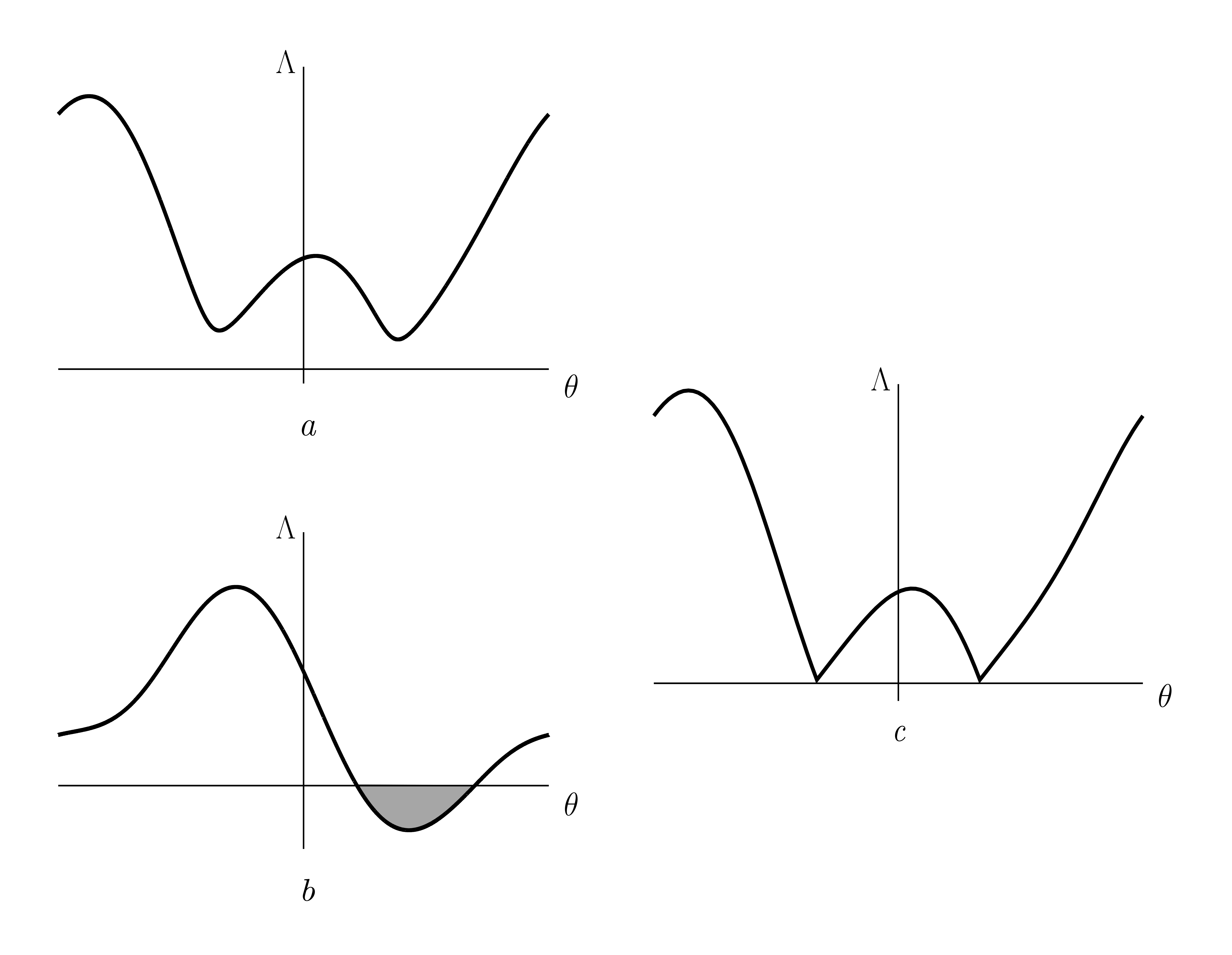}} 
    \caption{Dispersion relation $\Lambda(\theta)$ for three representative 
             Hamiltonians that violate parity. In the case $a$ the theory has 
             a gap, hence its ground state is $\ket{0}$, which preserves 
             parity. The dispersion relation in $b$ is negative for a set 
             of modes (shadowed interval). Therefore the theory is gapless and
             the ground state is obtained by filling these 
             modes (the Dirac sea), breaking the parity of the state. 
             In the panel $c$, $\Lambda$ is non-negative but it has zeros. 
             Then the model is gapless but the ground state is
             $\ket{0}$ and, hence, invariant under parity.}
  \label{logar_block}
   \end{figure}

A general stationary state is obtained by selecting a set of modes
$\mathbf{K}\subset \{0, \dots, N-1\}$ to which we associate
 \begin{equation}\label{eigenstates}
  \ket{\mathbf{K}}=\left(\prod_{k\in\mathbf{K}}d_k^\dagger\right)\ket{0}, \qquad 
{\rm with} \qquad d_k|0\rangle =0, \quad \forall k,
 \end{equation}
so that $\ket{0}$ represents the vacuum of the Fock space for the Bogoliubov 
modes $d_k$. The corresponding energy eigenvalue is
 $$E_{\mathbf{K}}= {\cal E}+
 \frac12\sum_{k\in\mathbf{K}}\Lambda(\theta_k) 
 - \frac12\sum_{k\not\in\mathbf{K}}\Lambda(\theta_k).$$
The ground state is obtained when we select the modes in the Dirac sea
 $$\ket{\rm GS}=\left(\prod_{\Lambda(\theta_k)<0} d_k^\dagger\right)\ket{0},$$
and has an energy
 $$E_{\rm GS}= {\cal E}-
\frac12\sum_{k=0}^{N-1}|\Lambda(\theta_k)|.$$
Notice that if
\begin{equation}\label{cond_nocrit}
  \Theta^-({\rm e}^{i\theta})^2\leq
  \Theta^+(\mathrm{e}^{i\theta})^2
  +|\Xi(\mathrm{e}^{i\theta})|^2,
\end{equation}
for all $\theta\in(-\pi,\pi]$, then the ground state is the Fock space vacuum 
$|0\rangle$; while if the inequality is not true for some values of $\theta$ then the 
state of minimum energy is obtained by exciting the modes with negative energy (Dirac 
sea). In Fig. \ref{logar_block} we represent these different possibilities. In the 
same figure, we also show a third possibility. This case corresponds to a critical 
gapless Hamiltonian with non-negative dispersion relation but vanishing at some values of 
$\theta$. It corresponds to situations in which (\ref{cond_nocrit})  holds but the 
right hand side vanishes at some points. As we shall see later the theory has 
different properties in the three cases and it requires a separate analysis.

It is interesting to discuss how the discrete transformations of
parity ($P$) and charge conjugation ($C$) act on this system.
In terms of the creation-annihilation operators they are defined by
\begin{equation}\label{parity-charge-def}
P:a_n\mapsto ia_{N-n},\quad C:a_n\mapsto a_n^\dagger.
\end{equation}
The Hamiltonian is $P$ invariant if the couplings $A_l$ are
real, while $PC$ is a symmetry if $B_l\in\mathbb{R}$. In the latter case,  
$H-\mathcal{E}$ goes to $\mathcal{E}-H$.

The action of $P$ and $C$ on the Fourier and Bogoliubov
modes are
\begin{eqnarray}
  &P:b_k\mapsto ib_{-k},\quad &C:b_k\mapsto b_{-k}^\dagger,
  \\
  &P:d_k\mapsto id_{-k},\quad &C:d_k\mapsto \overline d_{-k}^\dagger,
\end{eqnarray}
where $\overline d_k=\cos\xi_k b_k - i \sin\xi_k b^\dagger_{-k}$ represents
the Bogoliubov mode  for the Hamiltonian transformed under $PC$, i.e.
replacing $B_l$ by its complex conjugate $\overline B_l$. Note that the Bogoliubov modes 
transform covariantly under parity even if the Hamiltonian is not invariant, while they 
are covariant under charge conjugation only if the Hamiltonian is $PC$ symmetric.

The above observations have important consequences for determining
the symmetries of the ground state. In particular, one can notice that if the Hamiltonian 
has a gap, the ground state is always $P$ invariant, irrespectively of the symmetries
of the Hamiltonian. This corresponds to plot $a$ in Fig. \ref{logar_block}.
On the contrary, if the dispersion relation attains negative values, the ground state is 
given by the excitation of the modes in the Dirac sea, and then it is not parity 
invariant. This situation is represented in the plot $b$ in Fig. \ref{logar_block}.
There is a third possibility plotted in Fig. \ref{logar_block} $c$, 
in which the Hamiltonian is gapless and the ground state is parity invariant.
The three scenarios will be analysed separately in the following sections.
As we will show, their ground state entanglement entropy behaves under M\"obius 
transformation in three different ways. 

In the following paragraphs we will introduce the basic definitions
and notation to discuss the R\'enyi entanglement entropy for this model.

We divide the chain into two subsystems $X\cup Y=\{1,\dots,N\}$, so 
that the Hilbert space factorizes as 
$\mathcal{H}=\mathcal{H}_X\otimes\mathcal{H}_Y$.
We are interested in the R\'enyi entanglement entropy $S_{\alpha, X}$
of the subsystem $X$ for the ground state $\ket{\mathrm{GS}}$ of (\ref{ham}).
If we introduce the {\it partition function} \cite{CarCal} 
\begin{equation}\label{partition}
  Z_{\alpha,X} = \Tr(\rho_X^\alpha),
\end{equation}
the entropy is defined as
\begin{equation}\label{entropy-2}
S_{\alpha,X}=\frac{1}{1-\alpha}\log Z_{\alpha,X}.
\end{equation}

Due to the properties of the ground state of a Gaussian theory, 
the Wick decomposition holds for the $n$-point correlation functions; 
and therefore $Z_{\alpha,X}$ and the entropy can be recast
in terms of the two-point correlations \cite{Latorre, Peschel},
\begin{equation}\label{Vcorr-1}
V_{nm}=\left< \left(
\begin{array}{c} a_n \\ a_n^\dagger 
\end{array}\right)~(a_m^\dagger\quad  a_m)\right>
-\delta_{nm}I, \quad n,m=1,\dots, N.
\end{equation}
In fact,
\begin{equation}\label{partition-2}
Z_{\alpha,X}=\det F_\alpha(V_X),
\end{equation}
where 
\begin{equation}\label{Fax-function}
F_\alpha(x)=\left(\frac{1+x}{2}\right)^\alpha+
\left(\frac{1-x}{2}\right)^\alpha
\end{equation}
and $V_X$ is the $2|X|$ by $2|X|$ dimensional block matrix $V_X=(V_{nm})$, 
$n, m\in X$ with $2\times2$ submatrices being
\begin{equation}\label{correlation}
  V_{nm}=\frac{1}{N}\sum_{k=0}^{N-1}{\mathcal G}(\theta_k) \mathrm{e}^{i\theta_k(n-m)},
\end{equation}
with
\begin{equation}\label{symbolGS}
  {\mathcal G}(\theta)
  =\left\{\begin{array}{rlcccc}
  I, &\quad \mbox{if} \quad \Lambda(\theta)<0\quad
  \mbox{and}\quad \Lambda(-\theta)>0,\\
 M(\theta), &\quad \mbox{if} \quad \Lambda(\theta)\geq0\quad
 \mbox{and}\quad \Lambda(-\theta)\geq0,\\
 -I, &\quad \mbox{if} \quad \Lambda(\theta)>0\quad
 \mbox{and}\quad \Lambda(-\theta)<0,\\
 \end{array}\right.
\end{equation}
and 
 $$M(\theta)=\frac{1}{\sqrt{\Theta^+({\rm e}^{i\theta})^2+|\Xi({\rm e^{i\theta}})|^2}}
 \left(\begin{array}{cc} 
 \Theta^+(\mathrm{e}^{i\theta}) & \Xi(\mathrm{e}^{i\theta}) \\
 \overline{\Xi(\mathrm{e}^{i\theta})} & -\Theta^+(\mathrm{e}^{i\theta})
 \end{array}\right).$$

In the following sections it will be crucial 
the analytic continuation $\mathcal{M}(z)$ of
$M(\theta)$ from the unit circle $\upgamma=\{z={\rm e}^{i\theta}\}$
to the Riemann sphere $\overline{\mathbb{C}}$, namely
\begin{equation}\label{symbol-M}
\mathcal{M}(z)= \frac{1}{\sqrt{\myPhi^+(z)^2-
\Xi(z)\overline{\Xi(\overline z)}}}
\left(\begin{array}{cc} \myPhi^+ (z) & \Xi(z) \\ 
-\overline{\Xi(\overline z)} & -\myPhi^+(z) \end{array}\right).
\end{equation}
Notice that we have used $\Xi(z^{-1})=-\Xi(z)$.

\begin{figure}[h]
  \centering
     \resizebox{16cm}{7cm}{\includegraphics{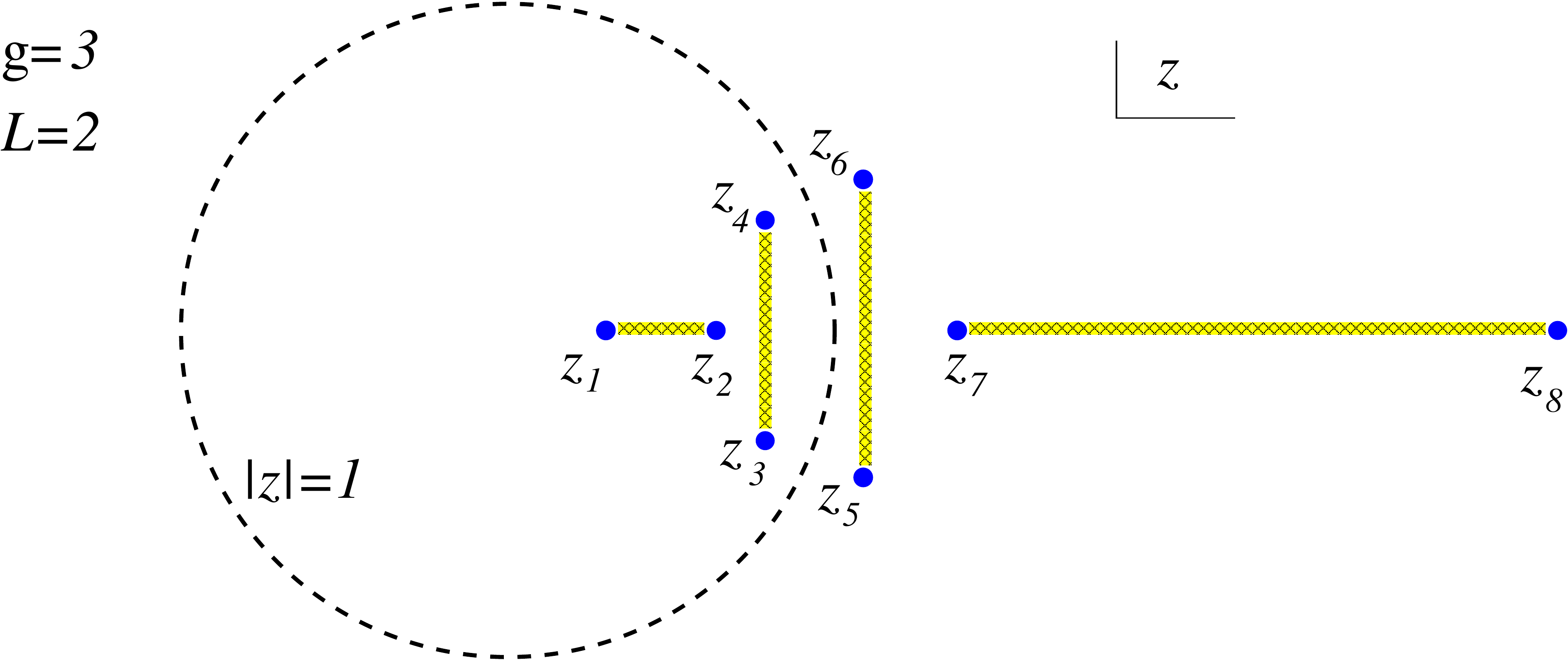}}
    \caption{Possible arrangement of the branch points and cuts 
             of $w=\sqrt{P(z)}$ for genus $\g=3$ ($L=2$). Half of
             them must be inside the unit circle and the rest outside
             it. Since the branch points must be related by inversion 
             and complex conjugation, $z_1=z_8^{-1}$, $z_2= z_7^{-1}$,  
             $z_3=\overline{z}_4=\overline z_5^{-1}=z_6^{-1}$.}
  \label{cuts}
   \end{figure}

To understand the analytic structure of $\mathcal{M}(z)$
we introduce the polynomial
\begin{equation}
P(z)=z^{2L}\Big(\myPhi^+(z)^2-\Xi(z)\overline{\Xi(\overline z)}\Big).
\end{equation}
Then, one immediately sees that $\mathcal{M}(z)$ is bivalued in $\overline {\mathbb C}$ 
and actually is meromorphic
in the Riemann surface represented by the complex curve
\begin{equation}\label{riemann-surface-1}
w^2=P(z).
\end{equation}
This curve represents a double covering of the Riemann sphere
with branch points at the zeros of $P(z)$. The genus of the Riemann surface is related to 
the range of the couplings of the Hamiltonian, ${\rm g}=2L-1$.
Due to the properties of $\myPhi$ and $\Xi$, 
one deduces that $P(z)$ has real coefficients and verifies 
$z^{4L}P(z^{-1})=P(z)$.
This implies that the zeros of $P(z)$ come in quartets
related by inversion and complex conjugation,
except for the real ones that come in pairs
related by inversion. In Fig. \ref{cuts} we present an
example corresponding to $L=2$.

The non-critical theories, i. e. those which have a mass gap or equivalently 
$\Lambda(\theta)>0$, can be easily characterized in terms of the roots of $P(z)$. 
To see it, simply consider the relation $$\left( 
\Lambda(\theta)+\Lambda(-\theta)\right)^2=4|P(\e^{i\theta})|.$$
Hence if $\Lambda(\theta)>0$ for any $\theta$, 
$P(\e^{i\theta})\not=0$ and there are no roots at the unit circle.

In the other case, that is, if $P(z)$ has roots at the unit circle,
they are necessarily degenerate, the dispersion relation vanishes at some
points and it corresponds to a critical massless theory.

In the next sections we shall take $X$ to be 
a single interval in the chain composed of $|X|$ consecutive sites,
namely $X=\{1,\dots,|X|\}$. In this case $V_X$ is a block 
Toeplitz matrix of the type that has been studied in 
\cite{Widom, Its2}. The more general case of $X$ composed of several
intervals will be discussed later in section \ref{sec:several}.

\section{M\"obius invariance in non critical chains}
\label{sec:non_critical}
We will consider in first place the case of non critical 
Hamiltonians for which $\Lambda(\theta)>0$ for any $\theta$.
Therefore, $\mathcal{G}(\theta)=M(\theta)$. In addition, we
are interested in the thermodynamic limit $N\to\infty$. Hence 
the sum in (\ref{correlation}) is transformed into the integral 
 \begin{equation}\label{correlationcomlex}
   V_{nm}=\frac1{2\pi i}\oint_{\upgamma} z^{n-m}\mathcal{M}(z)\frac{dz}z,
 \end{equation}  
where $\mathcal{M}(z)$ is the analytic continuation of $M(\theta)$
from the unit circle $\upgamma$ to the Riemann sphere defined
in (\ref{symbol-M}).

In \cite{Ares4} we introduced a new symmetry of the R\'enyi entanglement 
entropy using the connection of the  latter with the solution
of a Wiener-Hopf problem. Here we present a more direct approach
to the subject based on the existence of a similarity
transformation for the correlation matrix in the asymptotic limit.

Following \cite{Its} we introduce the operator
\begin{equation}\label{K-operator}
K_X \v (z) = \v(z) -\frac1{2\pi i}\oint_{\upgamma}\frac{(z/y)^{|X|}-1}{z-y}
\left(I-\mathcal{M}(y)\right)\v(y)\,\mathrm{d} y,\quad  
\v\in L^2({\upgamma})\otimes\mathbb{C}^2,
\end{equation}
defined to act on the Hilbert space $L^2({\upgamma})\otimes\mathbb{C}^2$ with scalar 
product
$$(\v_1,\v_2)=\frac1{2\pi i}\oint_{\upgamma} 
  \v_1(y)^\dagger\v_2(y)\,\frac{\mathrm{d} y}y.$$
In the topology induced by this inner product, $K_X$ is a continuous operator for any 
$\mathcal{M}$ bounded in ${\upgamma}$ and, as we shall show now, it fulfills the 
important property
\begin{equation}\label{detdet}
\det F_\alpha(V_X)=\det F_\alpha(K_X).
\end{equation}
In the following, we prove this fact.

Consider an  orthonormal basis in  $L^2({\upgamma})\otimes\mathbb{C}^2$
of the form $\{z^n\mathbf{e}_\nu | n\in\mathbb{Z}, \nu=1,2\}$, where 
the vectors $\mathbf{e}_\nu$ form the standard basis in $\mathbb{C}^2$. 
One can immediately compute the matrix representation of 
$K_X$ in this basis to get
$$
(K_X)=\left(
\begin{array}{c|c|c}
  I&0&0\\
\hline
(V_{na})&(V_{nm})&(V_{nb})\\
\hline
0&0&I
\end{array}
 \right),
$$
where the different indices are meant to run through the following
ranges:
$a\le 0$, $n,m=1,\dots,|X|$ and  $b>|X|$.
Now due to the block form of $K_X$ one has
$$\det F_\alpha(K_X)= \det F_\alpha(I)^2\cdot\det F_\alpha(V_X),$$
and looking at the definition (\ref{Fax-function}) 
of $F_\alpha$ one immediately obtains $F_\alpha(I)=I$, 
from which the result follows.

The relation between $K_X$ and $V_X$ can be extended to the
$|X|\to\infty$ limit. 
We may define
$$K\v(z)=\v(z) + \lim_{\mu\to 1^-} 
\frac1{2\pi i}\oint_{\upgamma}
\frac{I-\mathcal{M}(y)}{\mu z -y}\v(y)\,\mathrm{d}y,
$$
where we take the lateral limit for real values of $\mu$ smaller than 1. 
The matrix representation of $K$ in the basis introduced before is
simply
$$
(K)=\left(
\begin{array}{c|c}
  I&0
  \\
\hline
(V_{na})&(V_{nm})
\end{array}
 \right),
$$
with $a\le 0$ and $n,m>0$.

If we call $V$ the inductive limit of $V_X$ when $|X|\to\infty$,
we can immediately extend the relation (\ref{detdet}) to this limit
and thus we have
\begin{equation}\label{detdet-2}
\det F_\alpha(V)=\det F_\alpha(K).
\end{equation}

We shall denote by $Z_\alpha$ and $S_\alpha$
the partition function and the entanglement
entropy in this limit $|X|\to \infty$,
$$Z_\alpha=\det F_\alpha(K),$$
$$S_\alpha=\frac{1}{1-\alpha}\log Z_\alpha.$$

The computation of the determinant of $K$ 
for real couplings $A_l, B_l$ has been the 
subject of references \cite{Its, Its2}.
There the authors transform this into a Wiener-Hopf
factorization problem. Later on we will use their results.
For the moment, however,  we are not interested into
the computation of the  partition function or the R\'enyi
entanglement entropy, but rather in its invariance properties.
Hence a simpler approach is enough.

We shall proceed in two steps. First we consider general
M\"obius transformations and check under which circumstances 
the determinant $Z_{\alpha}=\det F_\alpha(K)$ is left invariant. 
In the next step we will ask which of these transformations 
are physical. By physical we mean those transformations that 
can be implemented as a change in the coupling constants of the 
theory.

Let us consider a M\"obius transformation on 
$\overline{\mathbb{C}}$ defined by
\begin{equation}\label{transf}
z'=\frac{az+b}{cz+d}, \qquad \left(\begin{array}{cc} a & b
\\ c & d\end{array}\right)\in SL(2,\mathbb{C}).
\end{equation}
This induces a transformation on the symbol $\CM(z)$ such that 
$\CM'(z')=\CM(z)$ and consequently on the operator $K$, so that
$$K'\v(z')=\v(z') + \lim_{\mu\to 1^-} 
\frac1{2\pi i}\oint_{{\upgamma}}\frac{I-\mathcal{M'}(y')}{\mu z'-y'}
\v(y')\,\mathrm{d}y',
$$
where for convenience we use primed integration variable. 
Now we perform the change of variables
$y'(y)$ induced by (\ref{transf}),
\begin{equation}\label{Kprime}
K'\v(z')=\v(z') + \lim_{\mu\to 1^-} 
\frac1{2\pi i}\oint_{{\upgamma}'}\frac{I-\mathcal{M}(y)}{\mu z'-y'(y)}
\v(y'(y))\,
\frac{\partial y'}{\partial y}\,
\mathrm{d}y,
\end{equation}
where the relation $\CM'(y'(y))=\CM(y)$ was used and
$\upgamma'=\{y, |y'(y)|=1\}$. 
The crucial property is that for any M\"obius transformation 
$$z'(z)-y'(y)=\left(\frac{\partial z'}{\partial z}\right)^{1/2}
\left(\frac{\partial y'}{\partial y}\right)^{1/2}
(z-y).$$
Plugging this into (\ref{Kprime}) we have
\begin{equation}\label{Kprimetwo}
K'\v(z'(z))=\v(z'(z)) + \left(\frac{\partial z'}{\partial z}\right)^{-1/2}
\lim_{\mu\to 1^-} 
\frac1{2\pi i}\oint_{{\upgamma}}
\frac{I-\mathcal{M}(y)}{\mu z-y}
\left(\frac{\partial y'}{\partial y}\right)^{1/2}
\v(y'(y))\,
\mathrm{d}y.
\end{equation}
Here we have assumed that we can safely apply the 
Cauchy's integral theorem in order to deform ${\upgamma}'$ 
into ${\upgamma}$, which in particular demands that
$\mathcal{M}$ is analytic in a region in with both
curves are homotopic. Defining 
$$T\v(z)= \left(\frac{\partial z'}{\partial z}\right)^{1/2}\v(z'(z)),$$
we finally obtain
$$T K' \v=K T \v.$$
Therefore $K$ and $K'$ are related by a similarity 
transformation and all its spectral invariants coincide. In particular
$\det F_\alpha(K)=\det F_\alpha(K')$.

To close the previous discussion, some words of caution are in order. 
On the one hand, in order to be able to apply the Cauchy theorem we need 
not only that $\CM(y)$ is analytic, as we actually demanded,  
but also $\v(y'(y))$ should have the same property. 
On the other hand, in the definition of $T$ we implicitly 
use analytic continuation for determining
$T\v(z)$ for $z\in{\upgamma}$. This implies that we should
restrict ourselves to situations in which $\v$ and $K'\v$
are analytic in a region such that ${\upgamma}$ and its M\"obius
transformed are homotopic.
Finally, note that the analyticity condition on $\CM(z)$ can be simply stated 
by asking that for any $z_i$ root of $P(z)$ inside (outside)
the unit circle, its M\"obius transformed $z_i'$ has to be 
also inside (outside).

The previous difficulties do not arise if the M\"obius 
transformation preserves the unit 
circle. In this case it is of the form
\begin{equation}\label{Mobius-circle-pres}
z'=\frac{az+b}{\bar b z+\bar a},\quad |a|^2-|b|^2=1.
\end{equation}
Then $T$ is a bounded operator 
and the similarity relation between $K$ and $K'$ holds
for any $\v\in L^2({\upgamma})\otimes\mathbb{C}^2$.

This is also the case of physical interest. The reason is that the
infinitesimal M\"obius transformation associated with changes of 
the coupling constants of the theory must preserve the unit circle. 

Indeed, as it is discussed in \cite{Ares4}, the M\"obius transformation 
(\ref{transf})
acts on the couplings $\mathbf A=(A_L, A_{L-1},\dots,A_{-L})$
and  $\mathbf B=(B_L, B_{L-1},\dots,B_{-L})$
like the spin $L$ representation 
of $SL(2,\mathbb{C})$. One immediately sees that only a subset of M\"obius transformations
map the original set of couplings $\mathbf A, \mathbf B$ to another
set $\mathbf A', \mathbf B'$ that fulfils the conditions
for the hermiticity of the Hamiltonian and the antisymmetry
of $\Xi'$: $A'_{-l}=\overline A'_l$ and $B'_{-l}=-B'_l$. 
An equivalent, and more appropriate way of imposing these 
conditions is to demand that the roots of the new polynomial 
$$P'(z')=(cz+d)^{-4L}P(z)$$ 
come also in quartets: if $z'_i$ is a root of $P'$
then $\bar z'_i$ and $1/z'_i$ are also roots. A way of 
guaranteeing this is by selecting those M\"obius transformations 
that commute with complex conjugation and inversion. 
These correspond in particular to the 
group $SO(1,1)$ with transformations like
\begin{equation}\label{so11}
  z'=\frac{z\cosh\zeta+\sinh\zeta}{z\sinh\zeta+\cosh\zeta}.
\end{equation}
Therefore, due to the physical meaning of our symmetry we are 
led to consider the 1+1 Lorentz group which has the important 
property of preserving the 
unit circle. 

{\bf To summarize:} M\"obius transformations in the $z$ plane induce a
change in the coupling constants of the free fermionic chain.
They are physically admissible if they are like
those in (\ref{so11}). Moreover, if we deal with a non critical theory, then
\begin{equation}
Z'_\alpha=Z_\alpha\qquad {\rm and} \qquad S'_\alpha=S_\alpha,
\end{equation}
where these are respectively the partition function of (\ref{partition})
and the R\'enyi entanglement entropy (\ref{entropy-2}) 
for an interval in the infinite size limit. 
This is easily derived from the relations $K'=T^{-1}KT$, 
$Z_\alpha=\det F_\alpha(K)$ and $S_\alpha=\log Z_{\alpha}/(1-\alpha)$,
that hold for physically admissible coupling constants.

\section{M\"obius transformations in critical theories}
\label{sec:critical}

Our next goal is to extend the considerations in the previous section
to the case of massless critical theories.
In this case, the symbol is not holomorphic in the unit circle
and the previous results do not apply. In addition, we must distinguish 
between ground states that are parity invariant and those that break 
this symmetry.

\subsubsection*{Parity invariant Ground States}

First, let us consider massless theories with a ground
state invariant under parity symmetry. An example is depicted in Fig. \ref{logar_block} 
$c$. In the previous section we associated a Riemann surface
to the fermionic chain. The critical point is attained when two
branch points collide at the unit circle. This corresponds to
the pinching of the corresponding Riemann surface. Hence ${\cal M}(z)$ is not holomorphic 
in the unit circle and the entanglement entropy scales logarithmically with the
size of the subsystem, with a coefficient proportional to the number
of pinchings.

The difficulty in this case is the following. The logarithmic 
term is well known and does not change under M\"obius
transformations. However we do not know in general how to compute the 
non-universal constant which differs from one theory to another.
In order to deal with this problem and determine how the entropy
behaves under M\"obius transformations we follow two different strategies.
The first strategy is to consider special cases in which we have an explicit
expression for the entropy. In this case we can easily deduce its behaviour
under the transformations. The second strategy is to study the limit
to criticality starting from a non-critical theory like those 
considered in the previous section. 
From the results obtained with the two strategies 
we will be able to conjecture a simple 
transformation law for the entropy. 
We cannot prove this conjecture in general but
numerical checks leave no doubt that it is correct.  

As we mentioned before, in some special situations
we have a completely explicit expression for the 
asymptotic behaviour of the entanglement 
entropy. For example, when $A_l\in{\mathbb R}$ and $B_l=0$.
In this case one immediately sees that the symbol 
$\CM(z)$ (\ref{symbol-M}) at the unit circle 
is $\pm\sigma_z$. The discontinuities are the roots of $z^L\Theta(z)$ at the 
unit circle, which we will denote by ${u}_\kappa=\exp(i\theta_\kappa), ~\kappa=1,
\dots,R$ in anticlockwise order. Due to the parity invariance,
 $\Theta(z^{-1})=\Theta(z)$, the roots ${u}_\kappa$ come in pairs related by inversion.
We shall assume that all of them are simple roots. 
This implies that they are different from $\pm 1$.

From this particular form of the symbol, it is clear that the entropy $S_{\alpha, X}$
and the partition function $Z_{\alpha,X}$
may depend solely on the set of roots $\underline {u}=({u}_1,\dots,{u}_R)$. 
Based on \cite{AEF}
(see also \cite{AEFS}), we can write the asymptotic behavior
in the following convenient way
\begin{equation}\label{entropy-AEF}
S_{\alpha,X}(\underline {u})=
\frac{(\alpha+1)
  R}{12\alpha}\log |X|-
\frac{\alpha+1}{12\alpha}
\sum_{1\leq\kappa\not=\upsilon\leq R}(-1)^{\kappa+\upsilon}
\log ({u}_\kappa-{u}_\upsilon)
+R I_\alpha+\dots,
\end{equation}
where 
\begin{equation}\label{Ialpha}
 I_\alpha=\frac{1}{2\pi i(1-\alpha)}\int_{-1}^{1} \frac{{\rm d} \log 
F_\alpha(\lambda)}{{\rm d}\lambda}
 \log\left[\frac{\Gamma(1/2-i\omega(\lambda))}{\Gamma(1/2+i\omega(\lambda)}\right]{\rm 
d}\lambda,
\end{equation}
with $\Gamma(z)$ the Gamma function,
\begin{equation}\label{omega-lambda}
\omega(\lambda)=\frac{1}{2\pi}\log\left|\frac{\lambda-1}{\lambda+1}
\right|,
\end{equation}
and the dots represent terms that vanish in the
large $|X|$ limit.

Now applying any M\"obius transformation
\begin{equation}\label{mobius-transform3}
{u}'_\kappa-{u}'_\upsilon
=\left(\frac{\partial {u}'_\kappa}{\partial {u}_\kappa}
\frac{\partial {u}'_\upsilon}{\partial {u}_\upsilon}\right)^{1/2}
({u}_\kappa-{u}_\upsilon),
\end{equation}
to the entropy (\ref{entropy-AEF}), we obtain in the asymptotic limit that
\begin{equation}\label{transf_S}
S'_{\alpha,X}(\underline {u}')=S_{\alpha,X}(\underline {u})+
\frac{\alpha+1}{12\alpha}\sum_{\kappa=1}^{R}
\log\frac{\partial {u}'_\kappa}{\partial {u}_\kappa}.
\end{equation}
We conjecture that this transformation
also applies for a general critical Hamiltonian with parity preserving vacuum.
To further motivate the conjecture and gain a better understanding of its origin,
we will study the limit to criticality of a symmetric theory.

Let us consider a general non critical Hamiltonian with parity and charge conjugation 
symmetry, i.e. $A_l,B_l\in\mathbb{R}$ $\forall l$. 
Criticality in this case is achieved when pairs of roots of $P(z)$, say
$z_{j_\kappa}$
and 
$\overline z_{j_\kappa}^{-1}$ approach ${u}_\kappa$  at the unit circle.
Geometrically this corresponds to the pinching of some cycles in the
associated Riemann surface.
As the roots approach its limit the entropy diverges.
In the Appendix C we compute its behaviour. The result
can be written as
\begin{equation}\label{divergence_gen}
  S_\alpha=-\frac{\alpha+1}{12\alpha}\sum_{\kappa=1}^R 
\log|z_{j_\kappa}-\overline{z}_{j_\kappa}^{-1}|
+K_\alpha(\underline {u})+\dots,
\end{equation}
where the dots stand for contributions that vanish
in the limit $z_{j_\kappa}\to {u}_\kappa, \kappa=1,\dots, R$.
We have also assumed that ${u}_\kappa\neq {u}_\xi$, $\kappa\neq \xi$
and we have omitted the explicit dependence
on the non degenerate branch points. 

We are interested in studying the behaviour of $K_\alpha$
under an admissible M\" obius transformation.
For that purpose we can use the invariance of $S_\alpha$. Indeed, one has
$$S'_\alpha=-\frac{\alpha+1}{12\alpha}\sum_{\kappa=1}^R
\log|z'_{j_\kappa}-\overline{z}_{j_\kappa}'^{-1}|
+K'_\alpha(\underline {u}')+\dots,$$
but
$$z'_{j}-{\overline{z}'_{j}}^{-1}=\frac{\overline z'_j}{\overline z_j}
  \left|\frac{\partial z'_j}{\partial z_j}\right|
(z_{j}-\overline{z}_{j}^{-1}).
  $$
Here we have used that M\"obius transformations in $SO(1,1)$ commute
  with complex conjugation. As these transformations also commute with inversion
  and the points ${u}_\kappa$ lie in the unit circle, we have
  $$S'_\alpha=-\frac{\alpha+1}{12\alpha}
  \sum_{\kappa=1}^R
  \left(\log|z_{j_\kappa}-\overline{z}_{j_\kappa}^{-1}|+  \log\frac{\partial 
{u}'_\kappa}{\partial {u}_\kappa}\right)
+K'_\alpha(\underline {u}')+\dots.$$
And from the invariance of $S_\alpha$ one has
\begin{equation}\label{Kalpha-transformed-1}
K'_\alpha(\underline {u}')=K_\alpha(\underline {u})+
\frac{\alpha+1}{12\alpha}
\sum_{\kappa=1}^R\log\frac{\partial {u}'_\kappa}{\partial {u}_\kappa}.
\end{equation}

It may seem strange that the entropy diverges
at the critical points ($z_{j_\kappa}=\overline{z}_{j_\kappa}^{-1}$). The reason for 
that is simple. We took the asymptotic $|X|\to\infty$ limit in the non critical theory. 
This renders a finite entropy. At criticality due to the logarithmic
scaling with the size of the subsystem, it diverges.
In fact, we may have a finite entropy in the critical theory by restoring
the finite size interval $X$. In this case the entropy of the critical
theory up to terms that vanish in the asymptotic limit is \cite{Ares3}
\begin{equation}\label{entropy_pinching_unit}
S_{\alpha,X} (\underline {u})= \frac{(\alpha+1)R}{12\alpha}
\log |X| +C_\alpha(\underline {u}) + \cdots,
\end{equation}
where we have reintroduced the dependence on the degeneration points ${u}_\kappa$.
The expression for the constant term $C_\alpha(\underline {u})$ is not known in general.

It is very suggestive to compare the formula above with (\ref{divergence_gen}).
One sees that, up to constant terms, (\ref{entropy_pinching_unit})
is obtained from (\ref{divergence_gen}) by simply replacing
$|z_{j_\kappa}-\overline{z}_{j_\kappa}^{-1}|$ with $|X|^{-1}$. If we
assume that the finite corrections are invariant under M\"obius
transformations, or in other words
$$K'_\alpha(\underline {u}')-C'_\alpha(\underline {u}')=
K_\alpha(\underline {u})-C_\alpha(\underline {u}),$$
then one immediately obtains a result identical to (\ref{transf_S}), that is,
\begin{equation}\label{crit_entropy}
S'_{\alpha,X}(\underline {u}')=S_{\alpha,X}(\underline {u})+
\frac{\alpha+1}{12\alpha}\sum_{\kappa=1}^{R}
\log\frac{\partial {u}'_\kappa}{\partial {u}_\kappa}.
\end{equation}
For the partition function, we have
\begin{equation}
  \label{crit_partition_sym}
Z'_{\alpha,X}(\underline {u}')=\prod_{\kappa=1}^{R}
\left(\frac{\partial {u}'_\kappa}{\partial {u}_\kappa}\right)^{2\Delta_\alpha}
Z_{\alpha,X}(\underline {u}),
\end{equation}
with $\Delta_\alpha=(\alpha^{-1}-\alpha)/24$. Therefore $Z_{\alpha,X}$ transforms
like the product of homogeneous fields of dimension $2\Delta_\alpha$
inserted at the pinchings $u_\kappa$.  

\begin{figure}[H]
  \centering
     \resizebox{15cm}{10cm}{\includegraphics{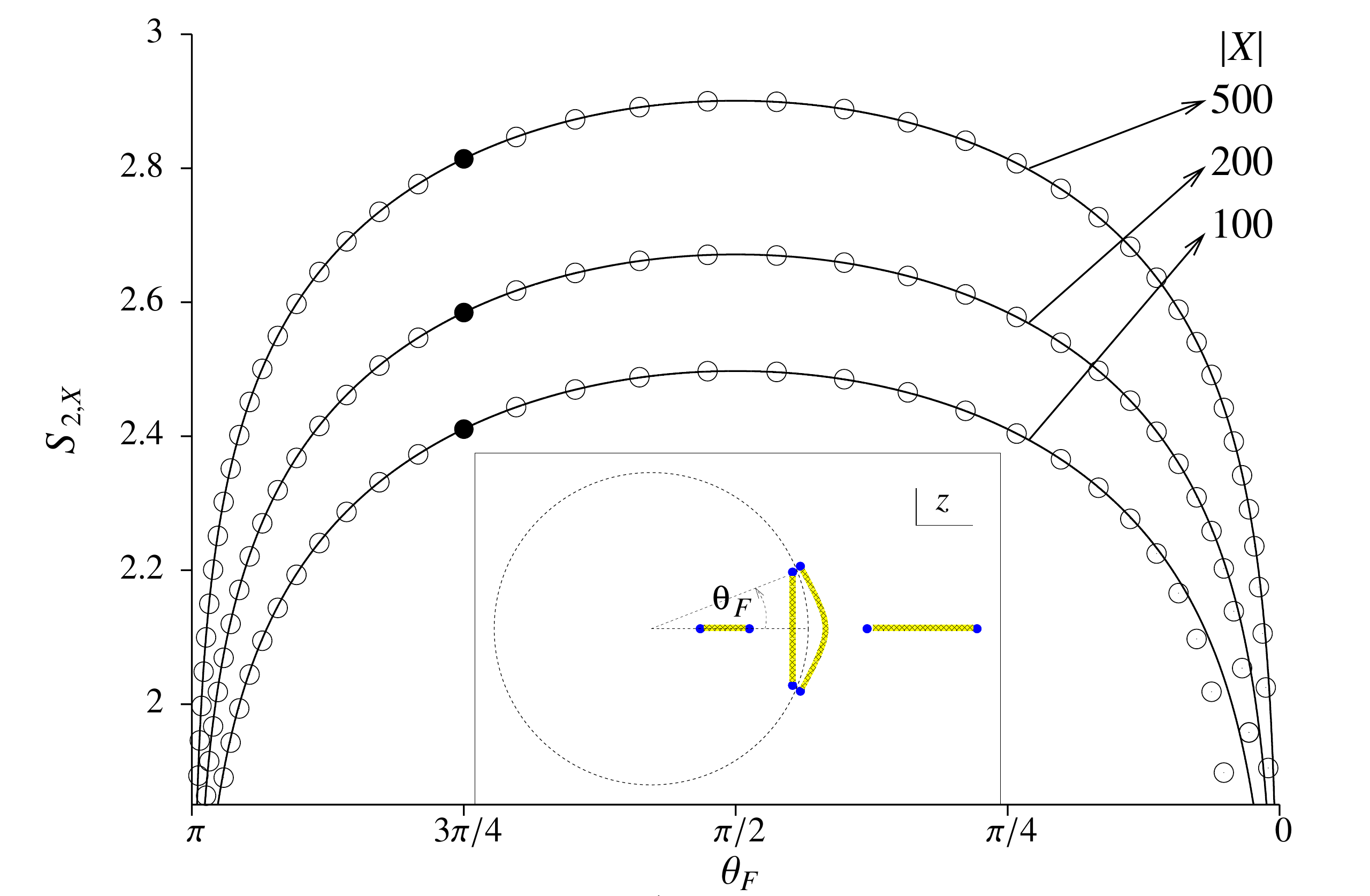}}
    \caption{Numerical check of transformation law (\ref{crit_entropy})
             for a critical theory with $L=2$ and pinchings at $\pm \theta_F$,
             as we depict in the inset. We have computed numerically for 
             several sizes of $X$ the change of the R\'enyi entaglement 
             entropy for $\alpha=2$ under $SO(1,1)$, (\ref{so11}).
             The solid line is our conjecture (\ref{crit_entropy}) 
             expressed in terms of the pinching angle $\theta_F$. The initial value of 
the entropy is set at $\theta_F=3\pi/4$ (filled dot). Observe that the finite 
size effects are relevant when $\theta_F$ approaches zero. In this case, all the 
roots of $P(z)$ are close to $z=1$, the stable fixed point of (\ref{so11}).}
  \label{numerics_pinchings}
   \end{figure}

We conjecture the same behaviour for the entanglement entropy
(or the partition function) of the free fermionic chain for any finite range critical 
Hamiltonian with parity symmetric vacuum. This has been checked
numerically. The results shown in Fig. \ref{numerics_pinchings} are quite convincing.

\subsubsection*{Ground states breaking parity symmetry}

The conjecture stated above was motivated by considering the critical theory as the limit 
of non critical ones. There are cases, however, in which this strategy is not possible.
Consider a theory with a non parity invariant vacuum
like the one plotted in Fig. \ref{logar_block} $b$. 
It is evident that this dispersion relation can not be considered 
as the limit of a non-critical theory. We shall therefore adopt 
a different strategy.

In the previous discussion we emphasized the close relation between
the logarithmic scaling of the entropy and the scaling dimension
of the partition function under M\"obius transformations. 
This scaling is associated to discontinuities of the symbol due to
the degeneration of branch points at the unit circle.
The idea is to apply the same relation with full generality.

As it was discussed in \cite{Ares3} (it is also easy to see from 
(\ref{symbolGS})), the discontinuities in the 
symbol for critical theories with vacuum preserving parity symmetry
correspond to a global change of sign. For cases in which the vacuum breaks parity 
invariance the discontinuities are of a different type. The symbol changes from $M$ to 
$-I$ at the Fermi point, $\theta_F$, or from $M$ to $I$ at its opposite value,
$-\theta_F$. 

The scaling behaviour of block Toeplitz determinants associated to matrix symbols with 
jump discontinuities, has been studied for the first time in \cite{Ares3}.
In that paper we propose an expression for the leading logarithmic scaling of the 
determinant when the two lateral limits commute.

The result is that the new discontinuities (with lateral limits  $M$ and $\pm I$) give 
a contribution to the coefficient of the logarithmic term of the entropy that is half of 
that associated to the pinchings at the unit circle (in which the two lateral
limits differ by a sign). Then, using the connection 
between the contribution to the logarithmic term and the scaling
dimension under admissible M\"obius transformations discussed above,
we propose that these new insertions have half the dimension of those
considered in the previous section. 

\begin{figure}[H]
  \centering
     \resizebox{15cm}{10cm}{\includegraphics{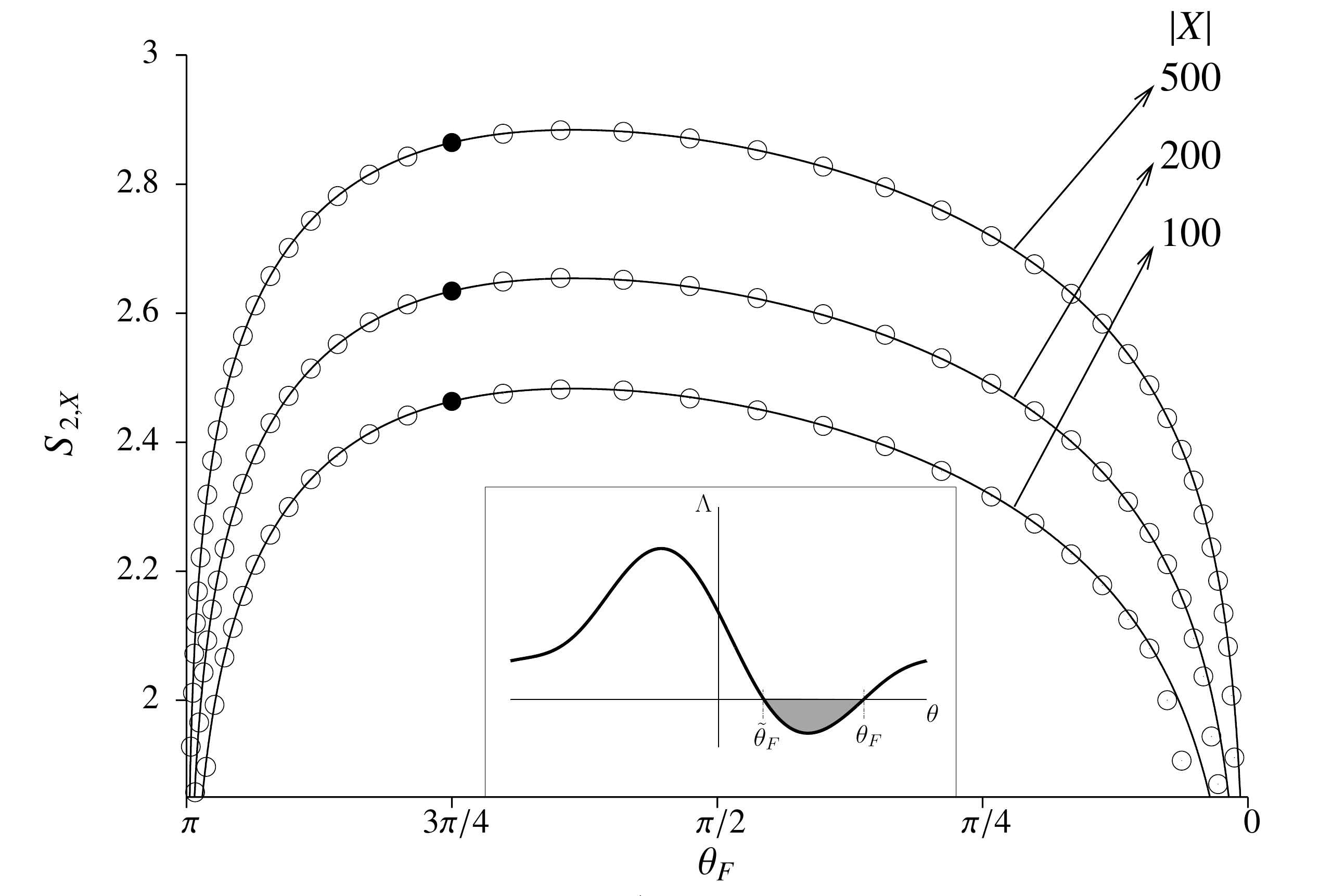}}
    \caption{Numerical check of the transformation (\ref{crit_partition}) 
             for a critical theory with $L=2$ and a Dirac sea depicted in 
             the inset. We plot the R\'enyi entanglement entropy with 
             $\alpha=2$ for different lengths of $X$ as a function of 
             the Fermi point $\theta_F$ under the $SO(1,1)$ group, (\ref{so11}). 
             The solid lines represent our conjectured transformation 
             (\ref{crit_partition}) expressed in terms of $\theta_F$. 
             The initial value of the entropy is set at $\theta_F=3\pi/4$ 
             and $\tilde{\theta}_F=\pi/2$ (filled dots). Observe that the finite 
             size effects are relevant when $\theta_F$ approaches zero. 
             In this case, all the roots of $P(z)$ are close to $z=1$, 
             the stable fixed point of (\ref{so11}).}
  \label{numerics_broken_parity}
   \end{figure}

If we add to the pinchings $u_\kappa$ discussed previously 
the points of the unit circle where the new jump discontinuities
(from $M$ to $\pm I$) take place and call these by 
$v_\sigma=\exp(i\theta_\sigma),\ \sigma=1,\dots,Q$, the behaviour under 
admissible M\"obius transformations of the partition function reads
\begin{equation}
  \label{crit_partition}
Z'_{\alpha,X}(\underline {u}',\underline{v}')=\prod_{\kappa=1}^{R}
\left(\frac{\partial {u}'_\kappa}{\partial {u}_\kappa}\right)^{2\Delta_\alpha}
\prod_{\sigma=1}^{Q}\left(\frac{\partial {v}'_\sigma}{\partial 
{v}_\sigma}\right)^{\Delta_\alpha}
Z_{\alpha,X}(\underline {u},\underline{v}).
\end{equation}
This expression covers the most general form for the behaviour under
admissible M\"obius transformations of the entanglement entropy of an interval
in the ground state of a free, homogeneous fermionic chain. Its
numeric accuracy is tested in Fig. \ref{numerics_broken_parity}.

We may generalize this discussion to excited states
and to subsystems composed of several intervals.  The
latter will be discussed in a subsequent section. But before doing
that we shall apply the transformations to the simplest fermionic
chain with range $L=1$, that via Jordan-Wigner transform is mapped
into the XY spin chain.

\section{Application to the XY-model with a Dzyaloshinski-Moriya coupling}

In this section we shall apply the previous results to the
XY spin chain with a Dzyaloshinski-Moriya (DM) coupling that
breaks the parity symmetry of the model. 
In particular we will show how to compute
the scaling behaviour of the entanglement entropy for
the critical Ising universality class.

As we mentioned before, a Jordan-Wigner transform
maps the model into a free fermionic chain with
nearest neighbours couplings. The coupling constants with the 
standard parametrization of the theory are $A_1=\overline A_{-1}=1+is$,
 $A_0=-h$ and $B_1=-B_{-1}=\gamma$; $h$ represents the transversal
 magnetic field, the DM coupling $s$ is a drift term that
breaks spatial parity invariance and $\gamma$ is the spin space anisotropy parameter.

\begin{figure}[h]
  \centering
    \resizebox{14cm}{11cm}{\includegraphics{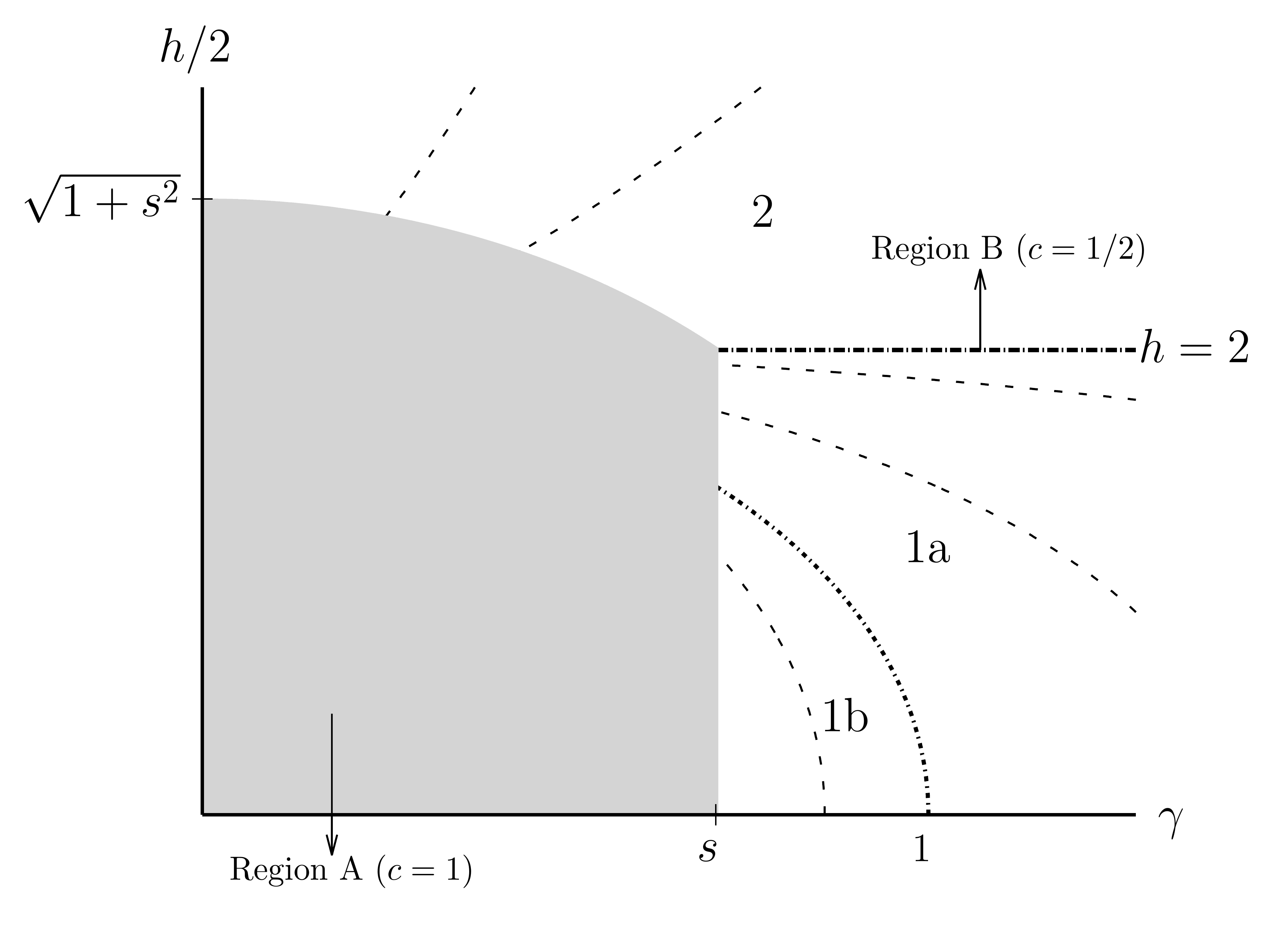}} 
    \caption{Phase diagram for XY spin chain with a DM coupling in 
            the $(\gamma, h)$ plane with fixed $s$. The shaded Region A 
            is gapless, belongs to the XX universality class and 
            its vacuum breaks parity. The dashed critical line labelled 
            as Region B  belongs to the Ising universality class and the 
            vacuum preserves parity. In the unshaded area, the Hamiltonian 
            has a gap and the flow of $SO(1,1)$ define conical curves of 
            constant entropy. 
            The dashed ellipses and hyperbolas depict some of them.}
  \label{phase_diagram}
   \end{figure}

The dispersion relation is
\begin{equation}\label{dispersion}
\Lambda(\theta)=
\sqrt{(h-2\cos \theta)^2+4\gamma^2\sin^2\theta}
+2s\sin\theta.\end{equation}
The theory is critical when
 $$ s^2-\gamma^2>0, \quad \hbox{and} 
 \quad \left(h/2\right)^2-s^2+\gamma^2<1; \quad\hbox{(Region A)},$$
 or when
 $$s^2-\gamma^2<0, \quad \mbox{and}\quad h=2;  \quad \hbox{(Region B)}.$$
In Fig. \ref{phase_diagram} we show a plot of the critical regions.
The ground state of the theories in Region A (for $s\not=0$) breaks 
parity invariance. Region B corresponds to the critical Ising universality line and the 
ground state is invariant under parity.

\subsubsection*{Parity preserving theories: $s=0$}

In the first part of this section we shall consider the parity
invariant Hamiltonian $s=0$, i.e. the pure XY spin chain. 
Moreover, $\gamma=0$ (no anisotropy) corresponds to the XX model and $\gamma=1$ to the 
quantum Ising model.

An interesting feature of the critical XX model ($s=\gamma=0, |h|<2$)
is that it is possible to compute analytically the full scaling
behaviour of the critical theory, including the finite term.
This is usually not known for a generic fermionic chain.
Using well know results based on the Fisher-Hartwig conjecture for Toeplitz
determinants we have \cite{Jin, AEFS, AEF}
\begin{equation}\label{critXX}
S_{\alpha,X}({u}_1,{u}_2)=\frac{\alpha+1}{6\alpha}\log |X| +
\frac{\alpha+1}{12\alpha}\log|{u}_1-{u}_2|+2I_\alpha+...,
\end{equation}
where ${u}_1=h/2+i\sqrt{1-(h/2)^2}$ and ${u}_2=\overline {u}_1$ are the two points at the 
unit circle in which the roots of $P(z)$ degenerate, and $I_\alpha$ defined in Eq. 
(\ref{Ialpha}) is the universal constant contribution.
 
Observe that the predicted transformation properties of the entropy
under admissible M\"obius transformations (\ref{crit_partition_sym}) 
are fulfilled, i.e.
$$
S'_{\alpha,X}({u}_1',{u}'_2)=S_{\alpha,X}({u}_1,{u}_2)+
\frac{\alpha+1}{12\alpha}
\log\left(\frac{\partial {u}'_1}{\partial {u}_1}
\frac{\partial {u}'_2}{\partial {u}_2}\right).
$$

Contrary to the case of the critical XX line, the scaling
behaviour for the entanglement entropy in the critical Ising
universality class ($h=2$) is only partially known. 
Until very short ago we were able to determine it
only for the special point that corresponds to the Ising model
($\gamma=1,h=2$). This was based on the relation
between the correlations of the Ising model and those of the XY model
without magnetic field after specializing it to their respective
critical points \cite{Igloi}, as well as by employing field theory methods
\cite{CardyCastro}. The result is
\begin{equation}\label{critI1}
  S_{\alpha,X}^{\gamma=1,h=2}=\frac12{S_{\alpha, 2X}(i, -i)}
  =\frac{\alpha+1}{12\alpha}\log |X| +
\frac{\alpha+1}{6\alpha}\log2+I_\alpha+...,
\end{equation}
where, for convenience, we have changed the notation
to make explicit the value of the couplings of the theory.
Recently an extension of this relation was proposed in \cite{Ares4}.
It allows to determine the entanglement entropy for
part of the critical line ($|\gamma|\leq1$). 
Here we present an alternative derivation of the same result,
based on the M\"obius transformation studied in the previous section.
An advantage of the new approach is that it can be applied to
the whole critical line.

We first consider the action of the  M\"obius transformation (\ref{so11})
on the coupling constants $h$ and $\gamma$ \cite{Ares4}, that is,
$$
\gamma'=\frac{\gamma}{h/2\sinh2\zeta+\cosh2\zeta},\qquad
h'/2=\frac{h/2\cosh2\zeta+\sinh2\zeta}{h/2\sinh2\zeta+\cosh2\zeta}.
$$
Observe also that in the critical Ising model two real
roots of $P(z)$ degenerate at $u=1$, hence
$$\left.\frac{\partial {u}'}{\partial u}\right|_{u=1}= {\rm e}^{-2\zeta}.$$
If we finally take $\gamma=\exp(-2\zeta)$ and apply (\ref{transf_S}),
we have
\begin{equation}\label{critIg-0}
S^{\gamma,h=2}_{\alpha,X}=S^{\gamma=1,h=2}_{\alpha,X}+
\frac{\alpha+1}{12\alpha}
\log\gamma.
\end{equation}
Alternatively, by using (\ref{critI1}), we obtain the simple result
\begin{equation}\label{critIg}
  S_{\alpha,X}^{\gamma,h=2}=\frac{\alpha+1}{12\alpha}\log 4\gamma|X| +
  I_\alpha+...
\end{equation}  
for the entanglement entropy along the critical Ising line $h=2$
and large $|X|$.

\subsubsection*{Parity breaking theories: $s\neq 0$}

In the second part of this section we consider $s\not=0$. This implies the breaking of 
parity invariance in the Hamiltonian.

As already discussed, if the theory has a mass gap the vacuum is still
parity invariant and the new coupling $s$ does not affect
the entanglement entropy. The same happens in the critical Ising
universality line. However in Region A ($ s^2-\gamma^2>0, \,\, \hbox{and}
\left(h/2\right)^2-s^2+\gamma^2<1$) the vacuum breaks parity invariance and the 
entanglement entropy does depend on $s$. The Fermi points $\theta_j$, $j=1, 2$ at which 
the dispersion relation changes sign, satisfy
\begin{equation}\label{fermi-points-Pbreak}
\cos \theta_j= \frac{-h/2\pm
  \sqrt{(s^2-\gamma^2)(s^2-\gamma^2+1-(h/2)^2)}}{s^2-\gamma^2+1},
\end{equation}
with $\theta_j\in(-\pi,0]$, for $s>0$. We have 4 discontinuities for the symbol
at the unit circle
$v_j={\rm e}^{i \theta_j}$
and
$v_{4-j}={\rm e}^{-i\theta_j}$, $j=1,2$.

In this region, an analytic expression for
the finite terms of the entropy is not known
except for some particular cases. Those will be discussed below. 
However, following the general discussion of the
previous section we can actually determine the 
behaviour of the finite terms under an admissible
M\"obius transformation.
First, notice that the new coupling constant transforms like
$$s'=\frac{s}{h/2\sinh2\zeta+\cosh2\zeta}.$$
Since the discontinuities of the symbol $\theta_j$ 
do not correspond to degenerate roots of $P(z)$, 
we have $R=0$, $Q=4$, and according to (\ref{crit_partition}),
\begin{equation}\label{XX-DM_transf}
  S^{\gamma',s',h'}_{\alpha,X}=S^{\gamma,s,h}_{\alpha,X}+
\frac{\alpha+1}{24\alpha}
\sum_{\sigma=1}^4\log \frac{\partial {v}'_\sigma}{\partial {v}_\sigma}.
\end{equation}

For the particular case $\gamma=0$ we can indeed compute the full
asymptotic expression for the entanglement entropy and check
the conjecture. In this situation the symbol is either $\pm I$ or $\pm\sigma_z$.
We can then reduce the problem to a scalar symbol where much more is
known (see e. g. \cite{AEFS}). Adapting the results from 
\cite{AEFS} to our situation we obtain
\begin{equation}\label{XX-DM_entropy}
S_{\alpha,X}=\frac{\alpha+1}{6\alpha}\log|X|+
\frac{\alpha+1}{12\alpha}
\log\left(4\frac{s^2-(h/2)^2+1}{s^2+1}\right)+2{I}_\alpha+\dots.
\end{equation}
This makes sense for $s^2-(h/2)^2+1>0$, i.e.
when we are in the critical region A with $\gamma=0$.
The case $h=2$ was already obtained in
\cite{Ares3}.

In order to verify the behaviour of (\ref{XX-DM_entropy}) 
under M\"obius transformations,
we need to compute the product of the complex Jacobians at the insertions $v_i$,
\begin{eqnarray}
\prod_{\sigma=1}^4\frac{\partial {v}'_j}{\partial {v}_j}
&=&
\left(\frac{s^2+1}{s^2+(h/2\sinh2\zeta+\cosh2\zeta)^2}\right)^2\cr
&=&
\left(\frac{s^2+1}{s'^2+1}\cdot\frac{{s'}^2-(h'/2)^2+1}{s^2-(h/2)^2+1}\right)^2.
\end{eqnarray}
From the second line one notice that (\ref{XX-DM_entropy}) transforms 
according to (\ref{XX-DM_transf}).

\section{Several disjoint intervals and the relation
with conformal invariance} \label{sec:several}

A different symmetry of the critical models is the conformal invariance.
This symmetry has been extensively studied in the context of
1+1 dimensional quantum field theory since it was recognized in
Ref. \cite{BPZ} as a powerful tool to determine the correlation functions of non trivial 
massless theories. These techniques have been applied in the context of entanglement 
entropy by different authors \cite{Holzhey, CarCal, CarCalTon}.

In order to compare the behaviour of the entropy under
M\"obius and conformal transformations, it is convenient
to extend the results of section \ref{sec:critical} 
to more general subsystems. As such we are going to consider here the case in which 
$X$ is composed of $P$ disjoint intervals, i.e.
\begin{equation}\label{disjoint-interval}
X=\bigcup_{i=1}^P [{x}_{2i-1}, {x}_{2i}].
\end{equation}
We also introduce a new notation for the partition function, i.e.
\begin{equation}\label{disj-int-partition}
Z_\alpha(\underline u,\underline v\;;\underline x)=\Tr(\rho^\alpha_X),
\end{equation}
where as before  $\rho_X$ is the reduced density matrix of the subsystem $X$
of the ground state of a critical theory, $\underline u=(u_1,\dots,u_R)$
are the positions of the pinchings at the unit circle
and  $\underline v=(v_1,\dots,v_Q)$ with $v_\sigma={\rm exp}(\theta_\sigma)$
are the points in the complex plane associated to the Fermi points or their opposite.

This case can not be analyzed with the tools used in the previous
sections, because the correlation matrix 
$V_X$ is no longer of the block Toeplitz type.
Rather, it is a principal submatrix of a block-Toeplitz matrix.

The correlation matrices associated to multi-intervals were studied in 
\cite{AEF}. From the results stated there, one derives in the asymptotic limit that
$$Z_\alpha(\underline u,\underline v\,;\underline x)=\prod_{1\leq\tau<\tau'\leq 2P}
Z_\alpha(\underline u,\underline v\,; x_\tau,x_{\tau'})^{-\sigma_\tau\sigma_{\tau'}},$$
where $\sigma_\tau=(-1)^\tau$. Observe that the partition functions in the
product of the right hand side correspond, all of them,
to those of a single interval $[x_\tau,x_{\tau'}]$ for different values of
$\tau$ and $\tau'$.
Therefore, the results of section \ref{sec:critical} apply.
Using (\ref{crit_partition}) and taking into account that
$$\sum_{1\leq\tau<\tau'\leq 2P}\sigma_\tau\sigma_{\tau'}=-P,$$
as can be easily deduced, one finally obtains
\begin{equation}
  \label{crit_multi}
  Z'_{\alpha}(\underline {u}',\underline {v}'
  \,;\underline {x})=
\prod_{\kappa=1}^{R}
  \left(\frac{\partial {u}'_\kappa}{\partial {u}_\kappa}\right)^{2P\Delta_\alpha}
  \prod_{\sigma=1}^{Q}
  \left(\frac{\partial {v}'_\sigma}{\partial {v}_\sigma}\right)^{P\Delta_\alpha}
Z_{\alpha}(\underline {u},\underline {v}\,;\underline{x})
\end{equation}
in the large $|x_\tau-x_{\tau'}|$ limit.

On the other hand, a global conformal transformation in space 
is given by
$$x'=\frac{ax+b}{cx+d}, \quad
\begin{pmatrix}a&b\\c&d\end{pmatrix}\in SL(2,\mathbb{R}),
$$
with $x,x'\in\mathbb{R}$.
  Under this transformation the partition function scales as
\begin{equation}
  \label{crit_conf}
  Z_\alpha(\underline u, \underline{v} \;;\underline x')=
\prod_{\tau=1}^{2P}
\left(\frac{\partial {x}'_\tau}{\partial {x}_\tau}\right)^{2C\Delta_\alpha}
Z_{\alpha}(\underline {u},\underline {v} \;;\underline{x},),
\end{equation}
where the central charge $C$ depends on the number of pinchings
$C=\frac{R}2+\frac{Q}4.$

It is manifest the striking similarity of this expression and
(\ref{crit_multi}) with the r\^ole of the endpoints of the intervals
$x_\tau$ replaced by the pinchings  $u_\kappa, v_\sigma$.

Actually, it is possible to obtain a unified expression if one consider
simultaneously M\"obius and conformal transformations.
Consider the map induced by an element of the direct product
$SO(1,1)\times SL(2,\mathbb R)$ on the space of couplings, pinchings and
endpoints
$$(\mathbf A, \mathbf B, \underline u, \underline v, \underline x)
\mapsto
(\mathbf A', \mathbf B', \underline u', \underline v', \underline x')$$
where the element in the first group factor acts as a M\"obius
transformation in $
\mathbf A, \mathbf B, \underline u$ and  $\underline v$
while the conformal transformations
in the second factor act on $\underline x$. The complex Jacobian determinant in
$(u_\kappa, x_\tau)$ or $(v_\sigma , x_\tau)$ is given respectively by
$$J_{\kappa\tau}=\frac{\partial {u}'_\kappa}{\partial {u}_\kappa}
\frac{\partial {x}'_\tau}{\partial {x}_\tau},$$
and
$$K_{\sigma\tau}=\frac{\partial {v}'_\sigma}{\partial {v}_\sigma}
\frac{\partial {x}'_\tau}{\partial {x}_\tau},$$
because the transformation of $u$ and $v$ do not depend on $x$ and viceversa.

\begin{figure}[H]
  \centering
     \resizebox{!}{8cm}{\includegraphics{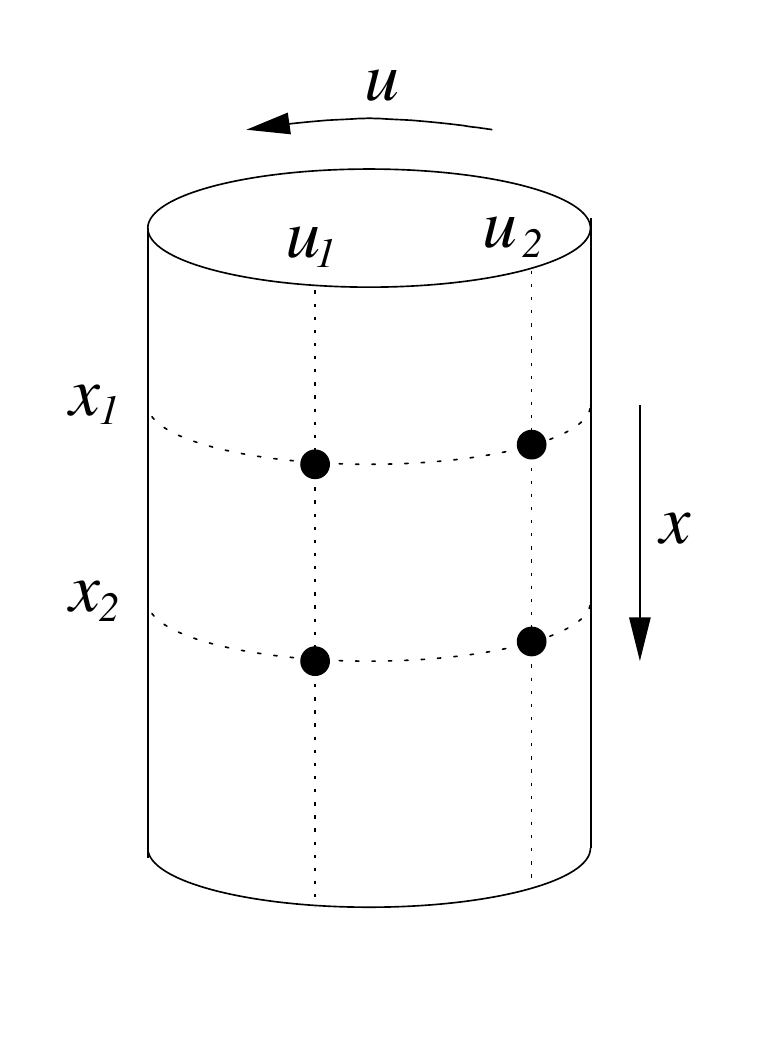}}
    \caption{The partition function $Z_{\alpha}$ of an interval $X=[x_1, x_2]$ 
             and pinchings at $u_1$, and $u_2$ behaves under 
             $SO(1,1)\times SL(2,\mathbb{R})$ as the expectation value 
             of a product of homogeneous fields inserted at the 
             points $(u_1, x_1)$, $(u_1, x_2)$, $(u_2, x_1)$ and $(u_2, x_2)$.}
  \label{cylinder}
   \end{figure}

Therefore, under a general transformation in $SO(1,1)\times SL(2,\mathbb R)$
eqs. (\ref{crit_multi}) and (\ref{crit_conf}) can be combined to give the
following transformation law
\begin{equation}
  \label{crit_unified}
  Z'_{\alpha}(\underline {u}',\underline v'\,;\underline {x}')=
  \prod_{\kappa,\tau} J_{\kappa\tau}^{\;\Delta_\alpha}
  \prod_{\sigma,\tau} K_{\sigma\tau}^{\;\Delta_\alpha/2}
  \ Z_{\alpha}(\underline {u},\underline v\,;\underline{x}).
\end{equation}
Note that this expression can be interpreted as the transformation
of the expectation value, in a covariant theory in $S^1\times{\mathbb R}$
of homogeneous fields of dimension $\Delta_\alpha$ at insertion points
$(u_\kappa, x_\tau)$ and of dimension $\Delta_\alpha/2$ at $(v_\sigma, x_\tau)$.
A particular example with $R=2$, $Q=0$, $P=1$ is depicted in Fig. \ref{cylinder} 

The natural question, whose answer we do not know yet,
is if there are more general
transformations and/or more general configurations
for the insertion points that lead to an expression
similar to (\ref{crit_unified}).

\section{Conclusions}

In this paper we have extended the results of Ref. \cite{Ares4}
on the invariance of the entanglement entropy under
M\"obius transformations. In that work we restricted ourselves to non-critical quadratic
spinless fermionic chains with parity and charge conjugation symmetries. In addition, we 
only considered subsystems formed by a single interval of contiguous sites. Here we have 
improved these results in several ways.

First, we present a more general and simple proof of the
above mentioned invariance. This allowed us to extend
the domain of application of the previous results
to include parity broken and/or charge symmetry broken Hamiltonians.

We then moved to the case of critical theories discussing how M\"obius 
transformations act on the corresponding {\it partition function}.
Critical theories are characterized by the existence
of jump discontinuities in the symbol, that is, the Fourier transformation of
the 2-point correlation function. We showed that in this case the partition function 
transforms like the product of homogeneous fields inserted at those discontinuities. 
The dimensions of these homogeneous fields depend on whether 
the discontinuity is associated to the existence or not of a Dirac sea. 
If there is a Dirac sea, then the ground state breaks parity symmetry. 
 
One striking aspect of this behaviour of the partition function is its parallelism
with the conformal transformations of primary fields in CFT. Notice though that the 
latter act on space-time itself. Actually, under conformal 
transformation the partition function behaves also
as the product of homogeneous fields but now inserted at 
the boundaries of the subsystem.
This relation is still more evident if we consider subsystems composed of several 
intervals. In this case, under M\"obius transformations, the insertions take place 
at the discontinuities of the symbol. 
Their dimensions depend on the number of end points of the 
intervals that form the subsystem. Conversely, under conformal transformations, the 
homogeneous fields are inserted at the end points of the intervals. Their dimensions 
depend on the number of discontinuities in the symbol.

This close similarity suggests that it is possible to consider both transformations 
together. This amounts to the introduction of a larger group that includes
M\"obius and conformal transformations. This larger group acts on a
{\it phase space} obtained from the cartesian product
of the space of discontinuities of the symbol (momentum space)
and of the end points of the intervals (real space). In this scenario the partition 
function transforms like the product of homogeneous fields inserted at the points 
consisting of combinations of discontinuities and end points.

The previous picture is very attractive. It immediately suggests 
a generalization of our results to larger groups containing the
direct product of M\"obius and conformal. Even more general configurations
in phase space can be envisaged.

Whether this project can be carried out, its meaning and further applications remains
an open question.

\noindent{\bf Acknowledgments:} Research partially supported by grants 
2016-E24/2, DGIID-DGA and FPA2015-65745-P, MINECO (Spain). 
FA is supported by FPI Grant No. C070/2014, DGIID-DGA/European Social Fund,
and thanks the hospitality and warm atmosphere of Instituto de F\'isica, Universidade
de Bras\'ilia where some of this research was carried out. 
ARQ is supported by CNPQ under process number 307124/2016-9 and thanks 
Departamento de F\'{\i}sica Te\'orica, Universidad de Zaragoza for the hospitality and nice 
atmosphere where part of this work was conducted.
FA and ARQ gratefully acknowledge support from the Simons Center for Geometry and Physics, 
Stony Brook University at which  some of the research for this paper was performed during 
the Program "Entanglement and Dynamical Systems". 
In particular, we thank Prof. V. Korepin for discussions during our stay in SCGP.

\appendix
\section{}

For completeness we shall review in this appendix the computation of the
entanglement entropy for a general parity and charge conjugation preserving
free fermionic Hamiltonian.
We follow \cite{Its, Its2, Ares4}.

\begin{figure}[h]
  \centering
    \resizebox{12cm}{4cm}{\includegraphics{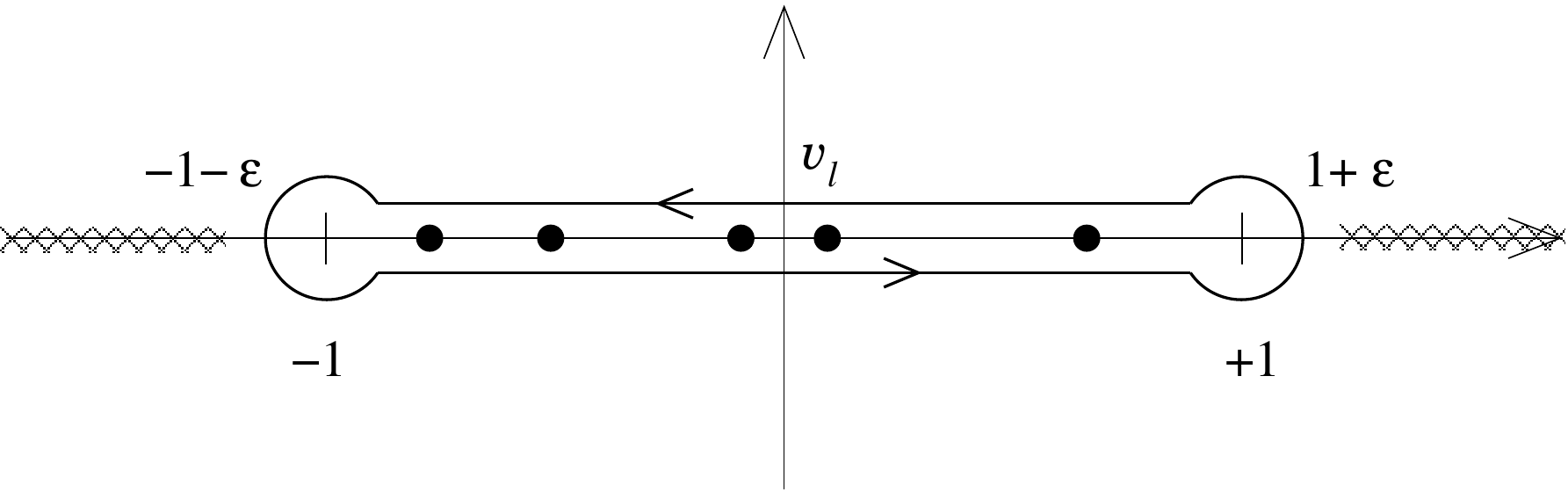}} 
    \caption{Contour of integration, cuts and poles for the computation of 
$S_{\alpha, X}$. The cuts for the function $f_\alpha$ extend to $\pm\infty$.}
  \label{contorno0}
   \end{figure}

It will be convenient to introduce the characteristic polynomial of
$V_X$, $D_X(\lambda)=\det(\lambda-V_X)$ in terms of which the entropy reads
\begin{eqnarray}\label{renyi1}
  S_{\alpha, X}=\lim_{\delta
\to 1^+}\frac{1}{4\pi i}\oint_{\mathcal{C}} 
f_\alpha(\lambda/\delta)\frac{\mathrm{d} \log
D_X(\lambda)}{\mathrm{d}\lambda}\mathrm{d}\lambda.
\end{eqnarray}
where $f_\alpha(x)= \log F_\alpha(x)/(1-\alpha)$ (see (\ref{Fax-function})) and 
$\mathcal{C}$ is the contour depicted in Fig. \ref{contorno0}
surrounding the eigenvalues $v_l$ of $V_X$,
all of them lying in the real interval $[-1, 1]$.

So far we have only transformed the problem of computing
the entropy into that of the characteristic polynomial of $V_X$.
But, it happens that an explicit expression  for $D_X(\lambda)$
is available if charge conjugation is preserved. In this case 
$A_l, B_l\in{\mathbb R}$ and the symbol reads
\begin{equation}\label{symbol-App}
\mathcal{M}(z)= U\left(\begin{array}{cc} 
0 & g(z) \\ g(z)^{-1} & 0 
\end{array} \right) U^{-1},
\end{equation}
with
$$U=\frac{1}{\sqrt{2}}\left(\begin{array}{cc} 1 & 1 \\
 -1 & 1 \end{array}\right),$$
and
\begin{equation}\label{g-function-App}
g(z)=\sqrt{\frac{\myPhi(z)+\Xi(z)}{\myPhi(z)-\Xi(z)}}.
\end{equation}

The roots of $P(z)= z^{2L}(\myPhi(z)+\Xi(z))(\myPhi(z)-\Xi(z))$
denoted by $z_j$ are either zeros or poles of the rational function $g^2(z)$.
According to this we assign an index $\epsilon_j$ to every root,
which is $+1$ if $z_j$ is a zero and $-1$ if it is a pole of $g^2(z)$.

Assume for the moment that all the roots of $P(z)$, denoted
by $z_1, z_2,\dots, z_{4L}$, are simple. The curve
$w^2=P(z)$ is a double covering of the Riemann sphere $\overline{\mathbb C}$
determining a genus $\g=2L-1$ Riemann surface.
Notice that we have $g^2(z)=\overline{g^2(\overline z)}=1/g^2(z^{-1})$. Hence the roots 
are related by inversion and conjugation, so that half of
them are inside the unit circle and the other half outside.
We order the roots such that $z_1,\dots,z_{2L}$ are inside the unit circle
and the rest lie outside. The cuts $\Sigma_\rho, \rho=0,\dots,{\rm g}$, in the complex 
plane are chosen to join $z_{2\rho+1}$ and $z_{2\rho+2}$ so that they do not 
intersect the unit circle. Adapted to our choice of cuts, we consider as a basis of 
the homology the cycles $a_r, b_r$ ($r=1,\dots,\g$) so that $a_r$ encloses the cut 
$\Sigma_r$ anticlockwise and the dual cycle $b_r$ surrounds the branch points $z_2, 
z_3\dots, z_{2r+1}$ clockwise. The canonical basis of holomorphic forms 
\begin{equation}\label{holomorphic-forms}
\mathrm{d}\omega_r
=\frac{\varphi_r(z)}{\sqrt{P(z)
}} \, { \rm d}z,
\end{equation}
with $\varphi_r(z)$ a polynomial of degree smaller than $\g$. It  
is chosen so that $\oint_{a_r} {\rm d}\omega_{r'}=\delta_{rr'}$.
The $\g\times\g$ symmetric matrix of periods $\Pi=(\Pi_{rr'})$ 
is defined by
\begin{equation}\label{periodic-matrix}
\Pi_{rr'}=\oint_{b_r} {\rm d}\omega_{r'}.
\end{equation}

We introduce the $\vartheta$ function with characteristics
$\chartheta{\vec p}{\vec q}
:{\mathbb C}^\g \rightarrow{\mathbb C}$ 
associated to $\Pi$ as
\begin{equation}\label{chartheta}
\chartheta{\vec{p}}{\vec{q}} (\vec{s}|\Pi)  
=\sum_{\vec{n}\in\mathbb{Z}^\g}
e^{\pi i (\vec{n}+\vec{p})\Pi\cdot(\vec{n}+\vec{p})
+2\pi i (\vec{s}+\vec{q})\cdot(\vec{n}+\vec{p})},
\quad \vec{p}, \vec{q}\in\mathbb{R}^\g.
\end{equation}
The \textit{normalized} version of the $\vartheta$ function is
\begin{equation}\label{charthera-norm}
\hattheta{\vec{p}}{\vec{q}}(\vec{s})= 
\hattheta{\vec{p}}{\vec{q}}(\vec{s}|\Pi)=
\frac{\chartheta{\vec{p}}{\vec{q}}(\vec{s}|\Pi)}
{\chartheta{\vec{p}}{\vec{q}}(0|\Pi)}.
\end{equation}

For large $|X|$ expansion, one may find that 
\begin{equation}\label{d_x_theta}
  \log D_X(\lambda)=|X|\log(\lambda^2-1)+\log\left(
  \hattheta{\vec\mu}{\vec\nu}(\beta(\lambda)\vec{e})\cdot
  \hattheta{\vec\mu}{\vec\nu}(-\beta(\lambda)\vec{e})
  \right)+\dots,
\end{equation}
where
$$\beta(\lambda)=\frac{1}{2\pi i}\log\frac{\lambda+1}{\lambda-1}$$ 
and $\vec{e}\in\mathbb{Z}^\g$ with its first $L-1$ entries equal to 0 and the last
$L$ equal to 1.
The characteristics $\vec{\mu}, \vec{\nu}\in(\mathbb{Z}/2)^\g$
depend on the indices assigned to the branch points and are defined by
$$\mu_r=\frac{1}{4}(\epsilon_{2r+1}+\epsilon_{2r+2});
\quad \nu_r=\frac{1}{4}\sum_{j=2}^{2r+1}\epsilon_j; \quad r=1, \dots, \g.$$
Recall that  $\epsilon_j=1$ if $z_j$ is a root of $g^2(z)$ (see (\ref{g-function-App}))
or $-1$ if it is a pole.

Of course, the expression for $D_X(\lambda)$ cannot depend
on the particular choice of cuts, provided none of them
crosses the unit circle. This has been
explicitly shown in \cite{Ares4}.
We shall make use of this fact further below.

It is also interesting to remark that the expression in
(\ref{d_x_theta}) is explicitly invariant under M\"obius 
transformations. Moreover, a M\"obius transformation move the branch points and cuts.
Thus the holomorphic forms (\ref{holomorphic-forms}) change, but not the matrix of
periods (\ref{periodic-matrix}). Therefore, the $\vartheta$ functions do not change
and then the entropy depending on them remains equal.

This property implies that $D_X(\lambda)$ only depends
on the set of independent cross ratios that can be obtained
with the branch points of the hyperelliptic curve. 
In addition, this symmetry is very useful to find dualities and 
other relations between different Hamiltonians in terms 
of the entanglement entropy. This has been studied in detail in Ref.
\cite{Ares4}.

\section{}

In this and the next Appendixes we shall study
the coalescence of pairs of roots of $P(z)$ degenerating
into a single double root. This produces a pinching of the associated 
complex curve $w^2=P(z)$. 

We will treat two separate cases. In one of them the roots degenerate at a
point at the unit circle.  In the other one they 
degenerate outside of the unit circle. In the first case the entropy diverges 
logarithmically while it has a finite limit in the second.
In this Appendix we will study the latter case. The first one will be treated in the 
next Appendix.

The analysis of a pinching of the hyperelliptic curve $w^2=P(z)$ 
requires the choice of a homology basis $(a, b)$. 
This will be conveniently chosen such that each pair of branch 
points being degenerated is surrounded by one of the $a$-cycles. 
An example of such basis is given in the Appendix A. 
In this basis, it becomes easy to deal with the 
divergences appearing in the matrix of periods. 
So, fix an order for the roots, compatible with the requirements 
of Appendix A. Let us focus on the situation 
when the roots $z_{j}$ and  $z_{j+1}$, with $j=2\hat r+1$, 
approach each other. In this case, the cycle $a_{\hat r}$ is enclosing them. 
Notice that this cannot be done if the two roots were in 
opposite sides of the unit circle. 

If we compute the period matrix we immediately see that the only 
entry that diverges when $z_j\to z_{j+1}$ is 
$\Pi_{\hat r\hat r}\sim -i/\pi \log|z_j-z_{j+1}|$. 
The remaining terms of this matrix, i.e. $\Pi_{l m}, l,m\not=\hat r$, 
are finite in this limit. 
Moreover, they define a new ${\rm g}-1\times{\rm g -1}$ period matrix denoted by 
$\Pi_\circ$. This new period matrix is associated to the Riemann surface resulting 
from the removal of the two merging branch points. The entries $\Pi_{l \hat r}, 
l\not=\hat r$ define a ${\rm g}-1$ dimensional vector denoted $\vec\Delta_\circ$.

There are two different cases for the resulting theta function after the limit. The 
simplest one is when the two degenerating points are of different character. This means 
that one is a zero and the other is a pole of $g^2(z)$. As a result $\mu_{\hat r}=0$. 
Due to the divergent behaviour of $\Pi_{\hat r\hat r}$, the only surviving terms in the
sum of (\ref{chartheta}) are those with $n_{\hat r}=0$. Hence, we have
\begin{equation}
  \lim_{
    z_j\to z_{j+1}
    }
  \chartheta{\vec {\mu}}{\vec 
{\nu}}(\vec{s}\,|\,\Pi)=\chartheta{\vec{\mu}_{\circ}}{\vec{\nu}_{\circ}}(\vec{s}_{\circ}
|\Pi_{\circ}),
\end{equation}
where $\vec{\mu}_{\circ}, \vec{\nu}_{\circ}$ and $\vec{s}_{\circ}$
stand for the ${\rm g}-1$ dimensional vectors obtained
from $\vec{\mu}, \vec{\nu}$ and $\vec{s}$, respectively, by removing the $\hat r$ entry.
Therefore, after the limit, we get a theta function associated to the
Riemann surface of genus ${\rm g}-1$ obtained after the removal of the 
two colliding branch points.

Before we continue with the other case, we may discuss the 
applications of this to  the entanglement entropy obtained 
from (\ref{d_x_theta}). We first recall that the branch points 
are related by inversion and conjugation. 
It means that if a pair of real roots merge there will 
be another merging pair related by inversion with the former.
In the complex case there will be three other coalescing pairs,
related by inversion and conjugation.
In both cases, all pairs are composed of roots of different 
character. Therefore, by successive application
of the coalescence limit we obtain
a $\vartheta$-function of genus either ${\rm g}-2$ for real roots
or ${\rm g}-4$ for complex ones. 

The result of the previous analysis is that the entanglement
entropy corresponds to that of a theory with range of couplings
either $L-1$ for the real case 
or $L-2$ for the complex one. It is interesting to observe that this
equivalence applies for the ground state entanglement entropy but not
for the dynamics: the respective Hamiltonians do not coincide.

Returning to the general discussion, the second case corresponds to two merging
roots that have the same character. Therefore, $\mu_{\hat r}=\pm1/2$. 
In the coalescence limit,  
$\chartheta{\vec \mu}{\vec \nu}$ vanishes. However, we are interested 
in the normalized theta function which does not vanish in this limit. 
We can compute its limit from
\begin{eqnarray}\label{pinching_outside}
\lim_{
    z_j\to z_{j+1}
}
\chartheta{\vec {\mu}}{\vec {\nu}}(\vec{s}\,|\,\Pi)\,
{\rm e}^{-\pi i\Pi_{\hat r\hat r}/4}
&=&
{\rm e}^{2\pi i (s_{\hat r}+\nu_{\hat r})\mu_{\hat r}} 
\,\chartheta{\vec{\mu}_{\circ}}{\vec{\nu}_{\circ}}(\vec{s}_{\circ}+\mu_{\hat r}
\vec{\Delta}_\circ|\Pi_{\circ})
\nonumber\\
&+& {\rm e}^{-2\pi i (s_{\hat r}+\nu_{\hat r})\mu_{\hat r}} 
\,\chartheta{\vec{\mu}_{\circ}}{\vec{\nu}_{\circ}}(\vec{s}_{\circ}-
\mu_{\hat r}\vec{\Delta}_\circ|\Pi_{\circ})\nonumber,
\end{eqnarray}
obtained from the non vanishing terms in (\ref{chartheta}), i. e.
those with $n_{\hat r}=0, -2\mu_{\hat r}$.

When applying this scenario to the entanglement entropy we observe 
that either two pairs of branch points, for the real case, or four
for the complex one, should merge simultaneously.
The resulting theta functions, 
after the merging, correspond to Riemann surfaces 
of genus either ${\rm g}-2$, for the real case, or ${\rm g}-4$. 
In contrast to the previous case of different character roots, 
in this limit the entanglement entropy is no longer equal to that
of a theory with a smaller range of couplings.

\section{}

Equipped with the previous results we can now study the behaviour of the
entropy for critical theories with parity symmetric vacuum, i.e.
those in which two branch points come together at the unit circle.
This corresponds to the pinching of a nontrivial cycle of the Riemann
surface determined by the symbol of the correlation matrix.

Here we are interested in the case of degenerating points of different type, i.e., 
one is a zero of $g^2(z)$ and the other is a pole. In this case, as it is discussed in 
Appendix B, the limit of the $\vartheta$ function is easily computed and coincides with 
that of a theory where the merging branch points are removed.

There is a problem, however, because the limiting procedure in the
previous Appendix was carried out assuming that one of the 
basic $a$ cycles of the homology encircles the merging points. 
However, the expression (\ref{d_x_theta}) for the characteristic 
polynomial of $V_X$ is valid when
no $a$ cycle intersects the unit circle. In this case, in which
the branch points degenerate at the unit circle, the two prescriptions are
not compatible.

\begin{figure}[h]
  \centering
     \resizebox{10cm}{!}{\includegraphics{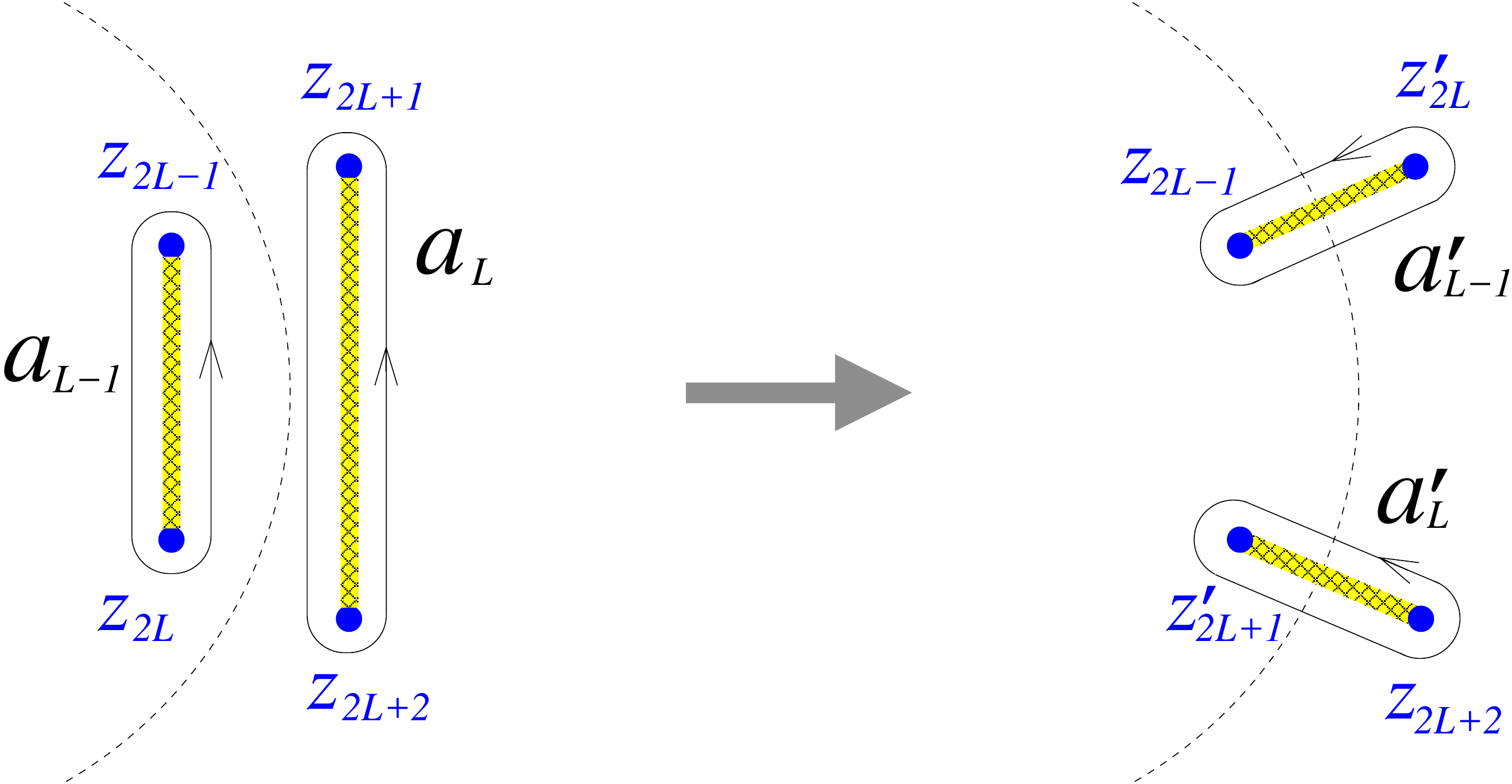}}
     \caption{Representation of the change of homology basis used
       to extract the divergent behaviour under a pinching. In the
       left panel the usual homology 
             basis described in Appendix 
             A, in which the $a$ cycles do not cross the unit circle,
             is depicted.
             In the right panel we represent the alternative homology
             basis in which $a'$ cycles enclose pairs of approaching
             branch points.
             This modular transformation is equivalent to a permutation
             in the labelling of the branch points: $z_{2L}=z'_{2L+1}$ 
             and  $z_{2L+1}=z'_{2L}$.}
  \label{modular}
   \end{figure}

The way to overcome this difficulty is by performing 
a modular transformation to a new basis $(a', b')$ 
where some $a'$-cycles cross the unit circle and enclose the pairs of
degenerating branch points. We shall initially order the roots so that
$z_{2L}$ degenerates with $z_{2L+2}=\overline z_{2L}^{-1}$ at the unit circle.
If the previous roots are real we do not impose other conditions to the
ordering. If they are not real we shall take $z_{2L-1}=\overline z_{2L}$
and  $z_{2L+1}=\overline z_{2L+2}$ that also degenerate at the unit circle.
In this ordering, where the first roots are inside and the last ones outside, a simple 
transposition of $z_{2L}$ and $z_{2L+1}$ induces the desired transformation of the basic
cycles, see Fig. \ref{modular}.

By recalling the results of \cite{Ares4} we obtain the following relation
between the original $\vartheta$-function and the new one,
\begin{equation}\label{pinching_unit}
\hattheta{\vec{\mu}}{\vec{\nu}}(\beta(\lambda)\vec{e}\,|\Pi)=
{\rm e}^{\pi i\beta(\lambda)^2 (\Pi'_{L-1,L-1}+\Pi'_{L,L}-2\Pi'_{L,L-1}-1)}
\hattheta{\vec{\mu'}}{\vec{\nu}}(\beta(\lambda)\vec{e'}\,|\Pi'),
\end{equation}
where $\Pi'$ is the period matrix for the new basic cycles,
$e'_r=e_r-(\Pi'_{L,r}-\Pi'_{L-1,r})$ and
$\vec{\mu'}$ differs from $\vec{\mu}$ only in
the $L-1$ and $L$ entries: 
$$\mu'_{L-1}=\mu_{L-1}+\nu_L-\nu_{L-1}+1/2,\qquad
\mu'_{L}=\mu_L-\nu_L+\nu_{L-1}+1/2.$$

We emphasize again that the advantage of using this basis of cycles
is that the divergences of $\Pi'$ are very simple to analyze.
In fact, for the real pinching (when $z_{2L}$ and $z_{2L+2}$ degenerate),
only $\Pi'_{L,L}$ diverges, so that
$$\Pi'_{L,L}\sim -i/\pi\,\log|z_{2L+2}-z_{2L}|,$$ 
while the rest of the entries of $\Pi'$ have a finite limit.
In the complex pinching (when also $z_{2L-1}=\overline z_{2L}$ and
$z_{2L+1}=\overline z_{2L+2}$ degenerate),
both $\Pi'_{L-1,L-1}$ and $\Pi'_{L,L}$ have a divergent behaviour,
$$\Pi'_{L-1,L-1}+\Pi'_{L,L}\sim -2i/\pi\,\log|z_{2L+2}-z_{2L}|.$$

Moreover, we should have $\epsilon_{2L}=-\epsilon_{2L+2}$
implying $\mu'_L=0$. In the complex case  we have also 
$\epsilon_{2L-1}=-\epsilon_{2L+1}$, so that $\mu'_{L-1}=\mu'_{L}=0$.
These observations are crucial to obtain the following finite limit
when the roots degenerate,
\begin{equation}\label{limit}
\hattheta{\vec{\mu'}}{\vec{\nu}}(\beta(\lambda)\vec{e'}\,|\Pi')
\ \longrightarrow\ 
\hattheta{\vec{\mu'}_\circ}{\vec{\nu}_\circ}(\beta(\lambda)\vec{e'}_\circ\,|\Pi'_\circ).
\end{equation}
Here on the right hand side we have the theta function for
a genus $\g-1$ Riemann surface in the real pinching or $\g-2$ in the complex one.
The period matrix $\Pi'_\circ$  is obtained in the real case by removing 
from $\Pi'$  the $L$ row and column. 
For the complex case, the removal are though the $L-1$ and $L$ rows and 
columns. Likewise, $\vec{\mu}_\circ$, $\vec{\nu}_\circ$ and $\vec{e'}_\circ$
stand for the vectors resulting after the removal of the $L$ component
if $z_{2L}\in{\mathbb R}$ or both the $L-1$ and $L$ ones if it is complex.

Taking into account the logarithmic divergence of $\Pi'_{LL}$ and eventualy
that of $\Pi'_{L-1,L-1}$ we finally obtain the following asymptotic
behaviour
\begin{equation}\label{asymptotic}
  \log\hattheta{\vec{\mu}}{\vec{\nu}}(\beta(\lambda)\vec{e}\,|\Pi)
 =-\frac c{2\pi^2}
  \left(\log\frac{\lambda+1}{\lambda-1}\right)^2
  \log|z_{2L}-z_{2L+2}|+\dots,
\end{equation}
where $c=1/2$ for the real pinching, $c=1$ for the complex one
and the dots stand for contributions which are finite in
the limit $z_{2L}\to z_{2L+2}$.

Finally, we plug (\ref{asymptotic}) into (\ref{d_x_theta}) and use
(\ref{renyi1}) to obtain 
\begin{equation}\label{divergence_uno}
S_\alpha\sim -c\frac{\alpha+1}{6\alpha}\log|z_{2L}-{z}_{2L+2}|.
\end{equation}

The above reasoning can be straightforwardly extended 
to situations where different couples of branch points
degenerate at the unit circle. For instance, $z_{j_\kappa}, 
\overline{z}_{j_\kappa}^{-1}\rightarrow {u}_\kappa$
with ${u}_\kappa=\exp(i\theta_\kappa)$, and 
${u}_\kappa\neq {u}_\xi$, $\kappa\neq \xi$. 
In this case, 
\begin{equation}
S_\alpha=-\frac{\alpha+1}{12\alpha}\sum_{\kappa=1}^R 
\log|z_{j_\kappa}-\overline{z}_{j_\kappa}^{-1}|
+K_\alpha(\underline {u})+\dots
\end{equation}
where the dots stand for contributions that vanish
in the limit $z_{j_\kappa}\to {u}_\kappa, \quad \kappa=1,\dots, R$
and $\underline {u}=({u}_1,\dots, {u}_R)$.
This is precisely the expression (\ref{divergence_gen}).

  \end{document}